\documentclass[aps,pra,reprint,groupedaddress,onecolumn,notitlepage]{revtex4-1}
\usepackage{color}
\usepackage{dcolumn}
\usepackage{amsmath}
\usepackage{amssymb}
\usepackage{mathtools}
\usepackage{graphicx}
\usepackage{latexsym}
\usepackage{amsfonts}
\usepackage{amssymb}
\usepackage{comment}
\usepackage[colorlinks=true,pdfstartview=FitV,linkcolor=black,citecolor=black,urlcolor=blue]{hyperref}

\newcommand{\be}{\begin{equation}}  
\newcommand{\ee}{\end{equation}}
\def\hc{\text{h.c.}}
\def\im{\mathrm{i}}
\def\ex{\mathrm{e}}
\def\ud{\mathrm{d}}
\def\VZ{V_{\text{Z}}}
\def\a{\alpha}
\def\b{\beta}
\def\g{\gamma}
\def\s{\sigma}
\def\w{\omega}

\begin{document}

%\preprint{}

\title{Electron correlation effects in superconducting nanowires in and out of equilibrium}


\author{Riku Tuovinen}
\email[]{riku.tuovinen@helsinki.fi}
%\thanks{}
\affiliation{QTF Centre of Excellence, Department of Physics, P.O. Box 43, 00014 University of Helsinki, Finland}


%\date{\today}

\begin{abstract}
One-dimensional nanowires with strong spin--orbit coupling and proximity-induced superconductivity are predicted to exhibit topological superconductivity with condensed-matter analogues to Majorana fermions. Here, the nonequilibrium Green's function approach with the generalized Kadanoff--Baym ansatz is employed to study the electron-correlation effects and their role in the topological superconducting phase in and out of equilibrium. Electron-correlation effects are found to affect the transient signatures regarding the zero-energy Majorana states, when the superconducting nanowire is subjected to external perturbations such as magnetic-field quenching, laser-pulse excitation, and coupling to biased normal-metal leads.
\end{abstract}

%\keywords{}

\maketitle

\section{Introduction}

One-dimensional nanowires may host Majorana zero modes (MZMs) when subjected to a suitable combination of spin--orbit interaction, proximity to an $s$-wave bulk superconductor, and an external magnetic field~\cite{Oreg2010,Lutchyn2010}. The MZMs' nonabelian statistics and their exponential localization at the opposite ends of the nanowire are highly desired properties for designing quantum computation with reduced decoherence issues due to topological protection~\cite{Kitaev2003,Nayak2008}. Even though the theoretical prescription is fairly simple, the experimental implementation for the observation of such topological signatures has proven extremely challenging~\cite{Mourik2012,Suominen2017,Prada2020,Yu2021}.

The Coulomb repulsion of electrons is present in any real material. While fairly large amount of work has been devoted to, e.g., disorder effects and how they affect the topological signatures~\cite{Lobos2012, Sau2013, Cole2016, Kaladzhyan2016, Andolina2017, Antipov2018, Liu2018, Thakurathi2018, Pan2020} (to mention just a few), also the interaction effects have received some attention~\cite{Gangadharaiah2011, Stoudenmire2011, Sela2011, Haim2014, Sticlet2014, Ruiz2015, Haim2016, Dominguez2017, Li2019, Wang2020, Vadimov2021}. In particular, for systems exhibiting the MZM, there is no guarantee of instantly relaxing to a steady-state configuration once the system has been driven out of equilibrium by applying an external perturbation~\cite{Weston2015, Francica2016, Bondyopadhaya2019, Tuovinen2019NJP, Vayrynen2020, Baranski2020}. To the best of the author's knowledge, only a few investigations of transient signatures of the MZM in the interacting case have been presented before~\cite{Vasseur2014, Wrzesniewski2021}, even in the clean limit. This is especially timely and relevant as state-of-the-art time-resolved pump--probe spectroscopy and transport measurements are pushing the temporal resolution down to the (sub-)picosecond regime~\cite{McIver2020, Lee2020, Nuske2020, Abdo2021, Budden2021, delaTorre2021}, where these effects could be observed in real time.

It is the purpose of this paper to study the electron-correlation effects and their role in the topological superconducting phase in and out of equilibrium. This investigation is carried out by using the nonequilibrium Green's function (NEGF) approach~\cite{Danielewicz1984, svlbook, Balzer2013, Schluenzen2020} within the generalized Kadanoff--Baym ansatz (GKBA)~\cite{Lipavsky1986, Spicka2021}. This approach allows for addressing both equilibrium and nonequilibrium properties at equal footing, and also for studying the interaction effects in a mathematically transparent and systematic way by the inclusion of the many-body self-energy.

After the Introduction, the paper is organized as follows. In Section~\ref{sec:model}, the model Hamiltonian for the superconducting nanowire is outlined. In Section~\ref{sec:model}, also the NEGF equations used for inferring both equilibrium and nonequilibrium properties are outlined. The equilibrium and nonequilibrium properties are studied in Sections~\ref{sec:eq} and~\ref{sec:noneq}, respectively. Diverse signatures of the transient build-up of the MZM are observed, depending on the electronic interaction, when the superconducting nanowire is subjected to external perturbations. Finally, Section~\ref{sec:concl} is a summary of the work, together with some prospects for future directions.

%%%%%%%%%%%%%%%%%%%%%%%%%%%%%%%%%%%%%%%%%%
\section{Model and method}\label{sec:model}

\begin{figure}[t]
\centering
\includegraphics[width=0.7\textwidth]{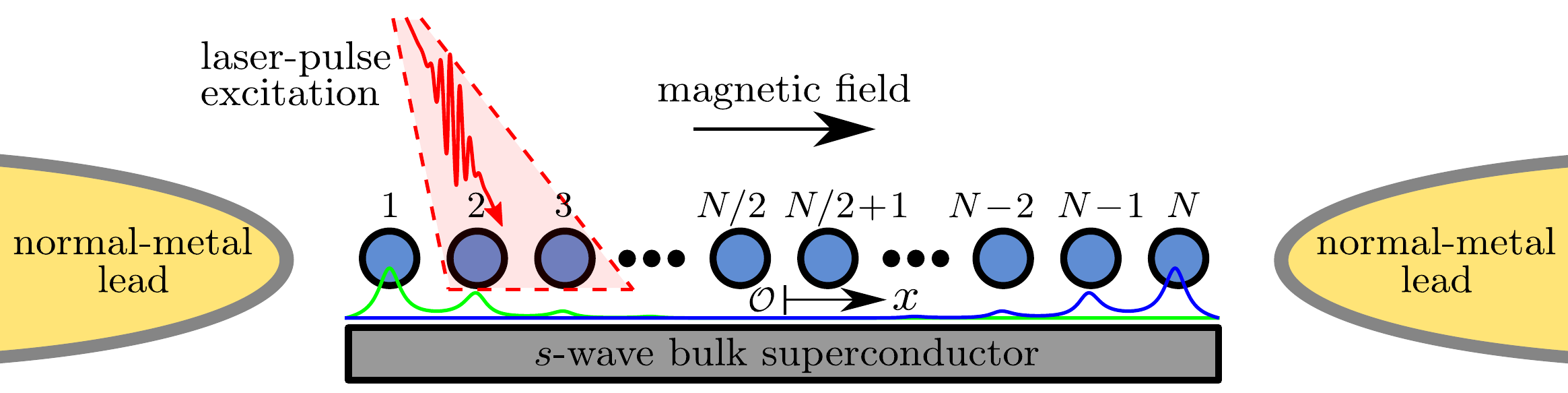}
\caption{Schematic of the studied model. A one-dimensional nanowire consisting of $N$ atomic sites (blue spheres) is in proximity to a bulk superconductor (grey slab) and in presence of a magnetic field. The origin ($\mathcal{O}$) of the real-space coordinate system ($x$) is set in the middle of the nanowire. The green and blue lines below the nanowire depict schematically the probability density of the two zero-energy states associated with the MZM, which are exponentially localized at the opposite ends of the wire. The nanowire is perturbed by an external laser pulse (red shaded area, see Section~\ref{sec:pulse}) and by contacting to normal-metal leads (Section~\ref{sec:transport}).}
\label{fig:setup}
\end{figure}

The studied system is a one-dimensional nanowire in proximity to an $s$-wave bulk superconductor, see Figure~\ref{fig:setup}. The nanowire is assumed to feature a strong spin--orbit interaction, for which suitable candidates include, e.g., InSb or InAs~\cite{Fasth2007,vanWeperen2015}. In addition, the nanowire is in the presence of an external magnetic field, which breaks the time-reversal invariance. The nanowire is characterized by the second-quantization Hamiltonian~\cite{Stoudenmire2011}
\begin{align}\label{eq:hamiltonian}
\hat{H} = \sum_i {\Big[} & -\frac{J}{2}(\hat{c}_i^\dagger \hat{c}_{i+1} + \hc) - (\mu-J)\hat{c}_i^\dagger \hat{c}_i - \frac{\a}{2}(\im\hat{c}_i^\dagger \s_y \hat{c}_{i+1} + \text{h.c.}) + \VZ \hat{c}_i^\dagger \s_z \hat{c}_i \nonumber\\
& + \varDelta(\hat{c}_{i\uparrow} \hat{c}_{i\downarrow} + \text{h.c.}) + U \hat{n}_{i\uparrow} \hat{n}_{i\downarrow} {\Big]} ,
\end{align}
where the sum runs over the nanowire sites $i\in[1,N]$, $J$ is the nearest-neighbor hopping between the nanowire sites, $\mu$ is the (equilibrium) chemical potential, $\a$ is the strength of the spin--orbit coupling, $\VZ$ is the Zeeman splitting (due to the external magnetic field), $\varDelta$ is the induced superconducting pairing potential, and $U$ is the on-site electron--electron repulsion (Hubbard type). The fermionic operators, $\hat{c}_{i\s}^{(\dagger)}$, annihilate (create) electrons from (to) site $i$ with spin orientation $\s\in\{\uparrow,\downarrow\}$. The density operator in the interaction term is defined as $\hat{n}_{i\s}=\hat{c}_{i\s}^\dagger \hat{c}_{i\s}$. The number of particles in the nanowire is determined by $\mu$. In Equation~\eqref{eq:hamiltonian}, the spin indices are summed over when suppressed, and $\s_y = \left(\begin{smallmatrix}0 & -\im \\ \im & 0 \end{smallmatrix}\right)$, $\s_z = \left(\begin{smallmatrix}1 & 0 \\ 0 & -1 \end{smallmatrix}\right)$ are the Pauli matrices. The external perturbations shown in Figure~\ref{fig:setup} for the laser pulse and the leads are discussed later, in Sections~\ref{sec:pulse} and~\ref{sec:transport}, respectively.

Dynamical properties of the system described by Equation~\eqref{eq:hamiltonian} are extracted by the one-particle Green's function~\cite{svlbook}
\be\label{eq:green}
G_{i\s,j\s'}(z,z') = -\im \langle \mathcal{T}_\g [\hat{c}_{i\s}(z)\hat{c}_{j\s'}^\dagger(z')] \rangle ,
\ee
which is an expectation value, with respect to the grand-canonical ensemble, of contour-ordered fermionic operators. These operators are represented in the Heisenberg picture, and the ensemble average can be expressed as a trace over the density matrix. The complex time variables $z,z'$ run over the Keldysh contour $\g \equiv (t_0,t) \oplus (t,t_0) \oplus (t_0,t_0-\im\b)$, where $t_0$ marks the beginning of an out-of-equilibrium process, $t$ is the observation time, and $\b$ is the inverse temperature. Expressed in the one-particle site basis of the nanowire, the Green's function matrix satisfies the equation of motion~\cite{svlbook}
\be\label{eq:eom}
[\im\partial_z - h(z)]G(z,z') = \delta(z,z') + \int_\g \ud \bar{z} \varSigma(z,\bar{z})G(\bar{z},z'),
\ee
where $h$ is the one-particle part of the Hamiltonian in Equation~\eqref{eq:hamiltonian} and $\varSigma$ is the the self-energy kernel. References~\cite{Danielewicz1984, svlbook, Balzer2013, Schluenzen2020} provide a thorough overview of the NEGF methodology.

In practice, the equation of motion~\eqref{eq:eom} is transformed into real-time Kadanoff--Baym equations by using the Langreth rules: Both the Green's function and the self-energy have components lesser ($<$), greater ($>$), retarded (R), advanced (A), left ($\lceil$), right ($\rceil$), and Matsubara (M) depending on their time coordinates on the contour~\cite{svlbook}. Concentrating on the equal-time limit on the real-time branch, $z = t_-, z'= t_+$, leads to
\be\label{eq:rho}
\frac{\ud}{\ud t} \rho(t) + \im [h_{\text{HF}}(t),\rho(t)] = -[\mathcal{I}(t) + \hc],
\ee
which is the equation of motion for the reduced one-particle density matrix $\rho(t) \equiv -\im G^<(t,t)$. In Equation~\eqref{eq:rho}, $h_{\text{HF}}(t) \equiv h(t) + \varSigma_{\text{HF}}(t)$ is the (time-local) Hartree--Fock (HF) Hamiltonian with
%$(\varSigma_{\text{HF}})_{i\s,j\s'}(t) = \delta_{ij}\delta_{\s\s'}U \rho_{i\s,i\s}(t)$ 
%\be
%(\varSigma_{\text{HF}})_{i\uparrow(\downarrow),j\uparrow(\downarrow)}(t) = -\im\delta_{ij}U G_{i\downarrow,i\uparrow}^<(t)
%\ee
\be
(\varSigma_{\text{HF}})_{i\downarrow(\uparrow),j\downarrow(\uparrow)}(t) = \delta_{ij}U \rho_{i\uparrow(\downarrow),i\uparrow(\downarrow)}(t)
\ee
for the Hubbard interaction~\cite{Schluenzen2020}. The time-nonlocal part in Equation~\eqref{eq:rho} is included in the collision integral
\be\label{eq:collint}
\mathcal{I}(t) = \int_{t_0}^t \ud \bar{t} [ \varSigma_c^>(t,\bar{t})G^<(\bar{t},t) - \varSigma_c^<(t,\bar{t})G^>(\bar{t},t) ] - \im \int_0^\b \ud \bar{\tau} \varSigma_c^\rceil (t,\bar{\tau}) G^\lceil(\bar{\tau},t),
\ee
where $\varSigma_c$ is the correlation self-energy. Here, this is approximated at the second-order Born (2B) level,
%$(\varSigma_c)_{i\s,j\s'}^\lessgtr(t,t') = U^2 G_{i\s,j\s'}^\lessgtr(t,t')G_{j\s',i\s}^\gtrless(t',t)G_{i\s,j\s'}^\lessgtr(t,t')$
\be
(\varSigma_c)_{i\downarrow(\uparrow),j\downarrow(\uparrow)}^\gtrless(t,t') = U^2 G_{i\downarrow(\uparrow),j\downarrow(\uparrow)}^\gtrless(t,t')G_{j\uparrow(\downarrow),i\uparrow(\downarrow)}^\lessgtr(t',t)G_{i\uparrow(\downarrow),j\uparrow(\downarrow)}^\gtrless(t,t') ,
\ee
for the Hubbard interaction~\cite{Schluenzen2020}. For a general Fermi--Hubbard model, the 2B approximation has been shown to be very accurate in the regime $U\lesssim J$, even compared to numerically exact methods~\cite{Hermanns2014,Lacroix2014}. It is worth noting that, because of the factor $1/2$ in front of the hopping term in Equation~\eqref{eq:hamiltonian}, an interaction strength $U$ in this modeling corresponds to $2U$ in the standard Fermi--Hubbard model literature. Therefore, in our setting, let us focus on the parameter regime $U/J\lesssim 0.5$. For larger interaction strengths, more accurate results could be achieved by, e.g., the $T$-matrix approximation~\cite{Schluenzen2017}.

A closed equation for the one-particle density matrix in Equation~\eqref{eq:rho} is obtained by using a reconstruction formula for the lesser/greater Green's functions by the GKBA
\be\label{eq:gkba}
G^\lessgtr(t,t') \approx \im [ G^{\text{R}}(t,t')G^\lessgtr(t',t') - G^\lessgtr(t,t)G^{\text{A}}(t,t') ],
\ee
and the propagators are approximated by their HF form
\be\label{eq:hf}
G^{\text{R}/\text{A}}(t,t') \approx \mp \im \theta [\pm(t-t')]\mathcal{T}\ex^{-\im\int_{t'}^t \ud \bar{t}h_{\text{HF}}(\bar{t})},
\ee
where $\mathcal{T}$ is the chronological time ordering. We shall discard the imaginary-time collision integral in Equation~\eqref{eq:collint} for now. This is a typical approach with the GKBA due to the lack of a GKBA-like expression for the mixed components $G^{\rceil,\lceil}$. The contribution therefore is obtained by the adiabatic preparation of the initial state, starting from the uncorrelated (or HF) system and turning on the many-body interaction in the 2B self-energy~\cite{Rios2011,Hermanns2012}. Equation~\eqref{eq:rho} is then solved numerically by a time-stepping procedure~\cite{Stan2009,svlbook,Balzer2013,Tuovinen2019pssb}.

%%%%%%%%%%%%%%%%%%%%%%%%%%%%%%%%%%%%%%%%%%
\section{Equilibrium properties}\label{sec:eq}

The nanowire hosts the MZM with, e.g., the following set of parameters: $J=1$, $\a=0.5$, $\VZ=0.25$, $\varDelta=0.1$, $\mu=0$, and when the nanowire is of length $N\geq 50$~\cite{Tuovinen2019NJP}. The hopping value therefore fixes the unit system. As the parameter space for this model is fairly large, let us concentrate on this representative point in the MZM regime and fix the length $N=50$ unless stated otherwise. It is worth mentioning that the size of the one-particle basis for the Green's function is $4N=200$ due to the spin$\otimes$particle--hole representation in Equation~\eqref{eq:hamiltonian}.

The equilibrium properties can be studied by evolving the system in time in the absence of external perturbations. In this context, it is useful to evaluate the momentum- and energy-resolved spectral function
\be\label{eq:kspec}
A(k,\w) = \frac{\im}{N} \sum_{ij} \ex^{\im k (x_i-x_j)}\int \ud \tau \ex^{\im \w \tau}[G_{ij}^>(T+\tau/2,T-\tau/2)-G_{ij}^<(T+\tau/2,T-\tau/2)],
\ee
where $x_i$ are the real-space lattice coordinates along the nanowire, $\tau\equiv t-t'$ is the relative-time coordinate, and $T\equiv (t+t')/2$ is the center-of-time coordinate. The real-space coordinate system is fixed with a lattice spacing of one, and the origin is set in the middle of the two center-most sites in the nanowire (see Figure~\ref{fig:setup}). Tracing out the $k$-resolved part gives the standard spectral function
\be\label{eq:spec}
A(\w) = \im\int \ud \tau \ex^{\im \w \tau}\text{Tr}[G^>(T+\tau/2,T-\tau/2)-G^<(T+\tau/2,T-\tau/2)].
\ee
When visualizing the spectral functions, it is useful to shift the frequency axis about the band center, $\w_c$, which is set between the two centermost eigenvalues of the HF Hamiltonian. It is worth noting that $\w_c$ depends on the interaction strength $U$, and it is not necessarily equal to $\mu$ which is simply a model parameter in this description [cf.~Equation~\eqref{eq:hamiltonian}]. Since $\mu$ fixes the filling of the system, $\w_c$ is also not necessarily in the middle of the highest occupied molecular orbital and the lowest unoccupied molecular orbital. Moreover, the electron--electron repulsion essentially gives rise to a charging energy and thus effectively renormalizes $\mu$~\cite{Stoudenmire2011}.

\subsection{Energy-band structure}

While the description in Equations~\eqref{eq:kspec} and~\eqref{eq:spec} will lack spectral information beyond the HF form in Equation~\eqref{eq:hf}, as the GKBA satisfies the condition, $G^{\text{R}}-G^{\text{A}}=G^>-G^<$, they are still useful for visualizing the energy-band structure.

\begin{figure}[t]
\centering
\includegraphics[width=0.7\textwidth]{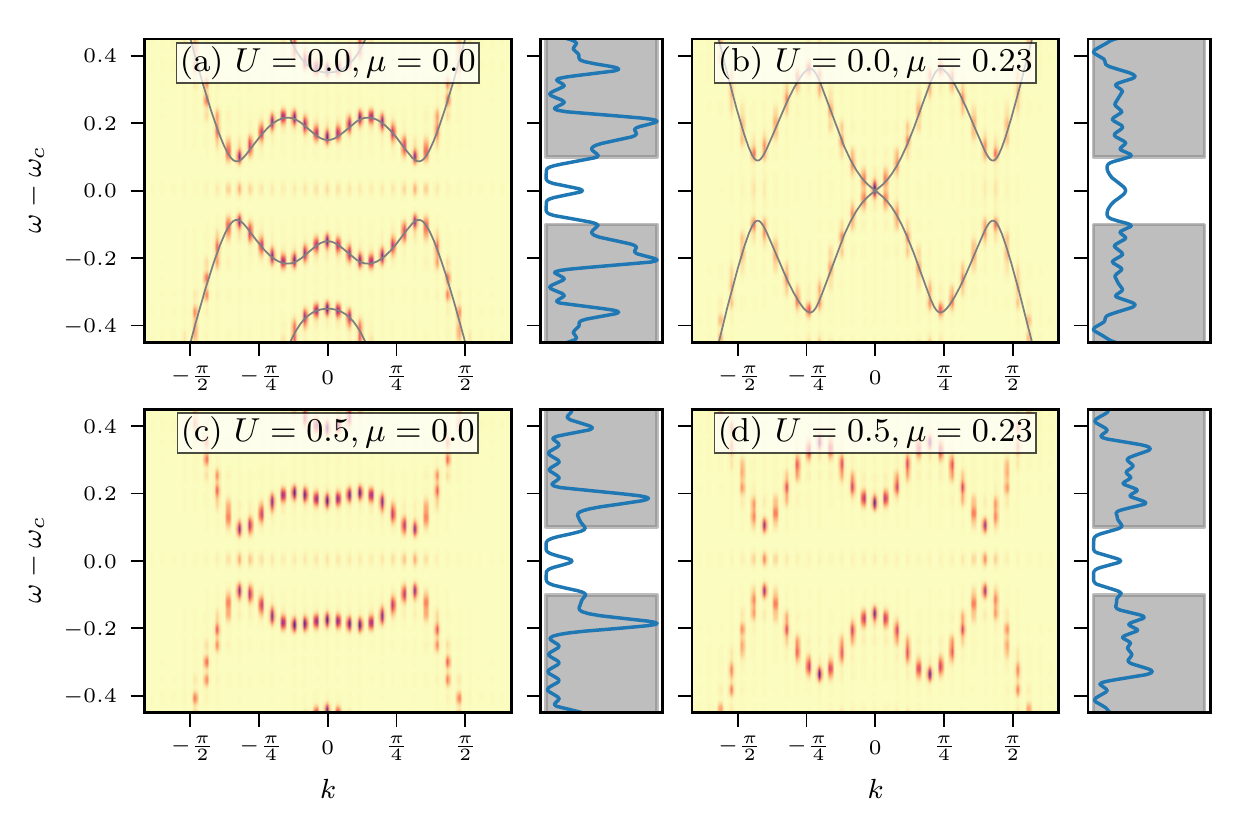}
\caption{Momentum and energy-resolved spectral function (color map; arbitrary units, where darker is higher) for the (a-b) noninteracting, $U=0.0$, and (c-d) interacting, $U=0.5$, nanowire. The cutouts on the right-hand side of the color maps show the $k$-integrated spectral function [Equation~\eqref{eq:spec}] with the vertical axes aligned with the color map. Analytically resolved energy bands for the infinite, noninteracting nanowire are superimposed with solid lines in panels (a-b) [Equation~\eqref{eq:hk}]. The shaded areas in the cutouts correspond to the states lying outside of the superconducting gap $|\w-\w_c|>\varDelta$. The fixed parameters are $J=1$, $\a=0.5$, $\VZ=0.25$, $\varDelta=0.1$.}
\label{fig:kspec}
\end{figure}

In Figure~\ref{fig:kspec}, the momentum- and energy-resolved spectral function [Equation~\eqref{eq:kspec}] for the nanowire is shown. This is evaluated by performing a time propagation up to $t=250J^{-1}$, and then taking $T$ at half the total propagation time. Then, the relative-time coordinate in Equation~\eqref{eq:kspec} spans the maximal range diagonally in the $(t,t')$ plane. The total bandwidth extends up to roughly $\pm 2J$ but let us concentrate on the low-energy states around the superconducting gap. In the noninteracting case, $U=0$, the energy bands for an infinitely long nanowire can be obtained analytically as the $k$-dependent eigenvalues of~\cite{svlbook}
\be\label{eq:hk}
h_k = a + b\ex^{-\im k} + b^\dagger \ex^{\im k},
\ee
where the on-site and nearest-neighbor contributions~\cite{Tuovinen2019NJP},
\be
a =
\begin{pmatrix}
J-\mu+\VZ & -\varDelta & 0 & 0 \\
-\varDelta & \mu-J+\VZ & 0 & 0 \\
0 & 0 & J-\mu-\VZ & \varDelta \\
0 & 0 & \varDelta & \mu-J-\VZ
\end{pmatrix},
\ee
\be
b =
\begin{pmatrix}
-J/2 & 0 & -\a/2 & 0 \\
0 & J/2 & 0 & -\a/2 \\
\a/2 & 0 & -J/2 & 0 \\
0 & \a/2 & 0 & J/2
\end{pmatrix},
\ee
respectively, are expressed in the spin$\otimes$particle--hole representation. The bands organize in the way of a standard topological superconductor. As the spin--orbit coupling breaks the spin-degeneracy, there are two sets of parabolas around $k=0$; these are further duplicated for particles and holes at positive and negative energies, respectively. The Zeeman splitting opens a gap at $k=0$, which makes it possible for the superconducting pairing potential to open another gap at $k\neq 0$, inducing a $p$-wave like pairing as long as~\cite{Oreg2010,Lutchyn2010}
\be\label{eq:toporegime}
\VZ^2>\mu^2+\varDelta^2 .
\ee
In this situation, the MZMs emerge in the case of a finite wire, as can be seen by the spectral peak at zero energy in Figure~\ref{fig:kspec}(a). In Figure~\ref{fig:kspec}(b), the situation corresponds to the boundary of Equation~\eqref{eq:toporegime}. At this point, the characterization changes qualitatively as the gap closes. For higher values of $\mu$, the gap would be opened again and the system would transform into an ordinary superconductor, without a peak at zero energy.

The MZMs are robust against the electron--electron interaction as the zero-energy peak remains for $U>0$, see Figure~\ref{fig:kspec}(c). This is in accordance with density matrix renormalization group (DMRG) data of Reference~\cite{Stoudenmire2011} where it was found that repulsive interactions enhance the effective Zeeman splitting while suppressing the pairing potential. Interestingly, we see for the elevated chemical potential in Figure~\ref{fig:kspec}(d) that the zero-energy state still pertains for the interacting system whereas the noninteracting system would already undergo a phase transition. While the topological phase is evidently protected against interactions, the parameter phase space may be extended beyond Equation~\eqref{eq:toporegime}. This shall be addressed next.

\subsection{Phase diagram}

\begin{figure}[t]
\centering
\includegraphics[width=0.7\textwidth]{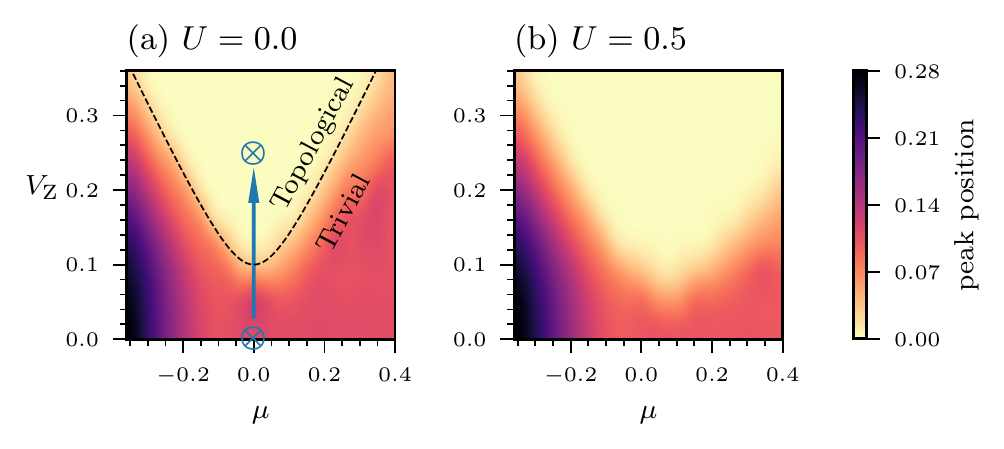}
\caption{Equilibrium phase diagram for (a) noninteracting, $U=0.0$, and (b) interacting, $U=0.5$, nanowire obtained by the low-energy spectral peak position (color map) for varying parameters $\VZ$ (vertical axes) and $\mu$ (horizontal axes). The dashed line in panel (a) shows the noninteracting phase boundary according to Equation~\eqref{eq:toporegime}. The crosses and the arrow in panel (a) indicate the quench calculation in Section~\ref{sec:quench}. The fixed parameters are $J=1$, $\a=0.5$, $\varDelta=0.1$.}
\label{fig:phase}
\end{figure}

Let us look at the spectrum more thoroughly. The position of the lowest-energy spectral peak is extracted for a wide range of parameters $\VZ$ and $\mu$ while keeping the other parameters fixed, and the equilibrium phase diagram is shown in Figure~\ref{fig:phase}. Here, the topological phase is attributed to the spectral peak position at zero energy. Again, in the noninteracting case the peak position adheres to the standard phase boundary according to Equation~\eqref{eq:toporegime}, see Figure~\ref{fig:phase}(a).

As already seen in Figure~\ref{fig:kspec}, the topological regime is extended to a larger chemical potential window on the right-hand side of Figure~\ref{fig:phase}(b). Enhancing the effective Zeeman splitting becomes more apparent when the filling ($\mu$) is increased, and thus, the many-body interactions become more significant. On the other hand, lower filling ($\mu$) on the left-hand side of Figure~\ref{fig:phase}(b) retains the noninteracting phase boundary as there are fewer interparticle interactions. Not only the effective Zeeman splitting is enhanced but also the effective pairing strength is suppressed due to interactions. The `minimum' of the phase boundary is lowered below $\VZ=\varDelta$ while it is also shifted to an elevated value for the chemical potential $\mu>0$. This confirms the robustness of the MZM against interactions and is in good agreement with the DMRG data of Reference~\cite{Stoudenmire2011}.

\section{Out-of-equilibrium dynamics}\label{sec:noneq}

As the NEGF+GKBA approach with the 2B self-energy has now been shown to capture the essential many-body effects of the interacting nanowire, let us then investigate how the interacting nanowire and the MZM are affected by external perturbations. Again, Equation~\eqref{eq:rho} is numerically evolved in time, and now an external perturbation is applied, driving the nanowire out of equilibrium.

\subsection{Sudden quench of magnetic field}\label{sec:quench}

\begin{figure}[t]
\centering
\includegraphics[width=0.7\textwidth]{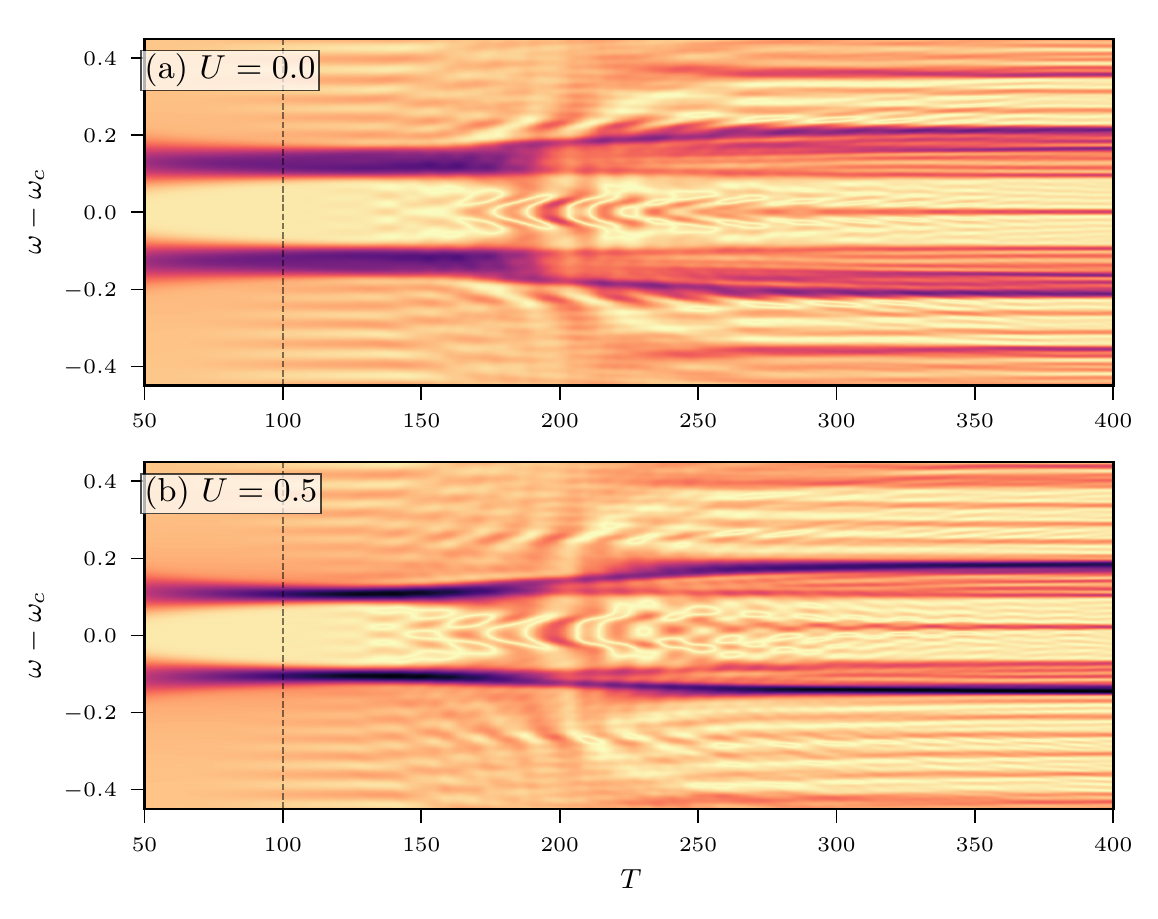}
\caption{Time-dependent nonequilibrium spectral function (color map; arbitrary units, where darker is higher) for (a) noninteracting, $U=0.0$, and (b) interacting, $U=0.5$, nanowire. The dashed vertical lines at $T=100$ indicate the instant of time when the system is quenched by suddenly increasing $\VZ=0 \to 0.25$. The fixed parameters are $J=1$, $\a=0.5$, $\varDelta=0.1$, $\mu=0$.}
\label{fig:quench}
\end{figure}

Figure~\ref{fig:quench} presents a calculation of the nonequilibrium spectral function, evaluated at different instances of the center-of-time coordinate $T$ [see Equation~\eqref{eq:spec}]. Also in this calculation, the center-of-time-coordinate is taken at $T=t/2$, where $t$ is the present instant in the time evolution. The beginning corresponds to the nanowire being in the ordinary superconducting phase ($\VZ=0$, $\varDelta=0.1$), and then the magnetic field is suddenly quenched ($\VZ=0.25$) at $t=200$ to drive the system towards the topological superconducting phase.
%The first signatures of the quench are therefore visible at the center-of-time-coordinate $T=100$.
The quench process is indicated by the arrow and crosses in Figure~\ref{fig:phase}(a).

Before applying the quench, the system is in the ordinary superconducting phase. Compared to the noninteracting case [Figure~\ref{fig:quench}(a)], the spectral peaks around the boundary of the superconducting gap at $|\w-\w_c|\approx\varDelta=0.1$ appear more bundled together for the interacting system [Figure~\ref{fig:quench}(b)]. At first, this seems counterintuitive as interactions typically broaden the spectral peaks. However, in our calculation, the spectral broadening is described at the HF level [cf.~Equation~\eqref{eq:hf}], so the effect here is more about multiple states being crammed together as the interaction is renormalizing the spectral peak positions [cf.~Figure~\ref{fig:kspec}(a,c)].

The quench is a very strong out-of-equilibrium condition, and the system behavior is crucially altered. In the noninteracting case [Figure~\ref{fig:quench}(a)], a spectral peak at zero energy starts forming after the quench. This spectral peak can, again, be attributed to the MZM. Interestingly, it takes from $T=100$ to $T=150$ (a duration of $50J^{-1}$) until the peak starts forming, which corresponds to the time for the information about the quench to cross the nanowire of length $N=50$. This build-up time of the zero-energy peak and its dependence on the nanowire length is crucial, since it corresponds to a pair state localized at the opposite ends of the nanowire. The amplitude of the zero-energy peak first oscillates and then saturates to a nonzero value as is consistent with the stationary state in Figure~\ref{fig:kspec}(a).

While the interacting case [Figure~\ref{fig:quench}(b)] retains the zero-energy peak in the stationary state, the initial transient oscillations are modified. The oscillation periods appear to be slightly longer compared to the noninteracting case. This is understandable as interparticle interaction introduces scattering events within the nanowire, thus obstructing the signal for reaching the nanowire ends, and the build-up time of the zero-energy peak becomes longer. This can be related to a similar effect of electron traversal times in nanojunctions being affected by disorder~\cite{Ridley2019entropy}. The frequency content of the transient oscillations is analyzed in more detail in Section~\ref{sec:pulse}. At around $T=250$, after multiple reflections of the wavepackets, higher energy side bands start to take shape. These correspond to the upper (unoccupied) and lower (occupied) bands in Figure~\ref{fig:kspec} split by the Zeeman energy. As this Zeeman splitting is effectively enhanced due to the interactions, the side peaks emerge further away from the main spectral peaks, when compared to the noninteracting case.

\subsection{Laser-pulse excitation and transient spectroscopy}\label{sec:pulse}

Let us then keep the nanowire characterized by the Hamiltonian in Equation~\eqref{eq:hamiltonian} with fixed parameters, and add an external laser-pulse excitation
\be\label{eq:pulse}
\hat{H}_{\text{ext}}(t) = \sum_{i\in I} E(t) \hat{c}_i^\dagger \hat{c}_i ,
\ee
where the pulse shape is taken as a gaussian, $E(t)=E_0 \sin(\varOmega (t-t_c)) \ex^{-4.6 (t-t_c)^2/t_c^2}$ of amplitude $E_0$, frequency $\varOmega$, and centering $t_c=2\pi n_c/\varOmega$ with $n_c$ being the number of optical cycles. In all calculations, $n_c=3$ is used. In Equation~\eqref{eq:pulse}, $I$ represents the set of atomic positions along the nanowire being irradiated by the pulse (see Figure~\ref{fig:setup}). Like in Equation~\eqref{eq:hamiltonian}, also here, the spin indices are summed over (suppressed), i.e., the pulse excitation is not spin selective.

To investigate how charge is (re-)distributed along the nanowire after the pulse excitation, let us look at the (field-induced) time-dependent dipole moment~\cite{Balzer2013}
\be\label{eq:dipole}
d(t) = \sum_i x_i \rho_{ii} (t) ,
\ee
where $x_i$ are, again, the real-space lattice coordinates along the nanowire, and $\rho_{ii}$ is the diagonal element of the density matrix, i.e., the site occupation number. The Fourier transform of the dipole moment, $d(\w) = \int \ud t \ex^{-\im \w t} d(t)$, gives us access to the spectral properties of the nanowire: The dipole spectrum $d(\w)$ is peaked at the excitation energies of dipole-allowed transitions~\cite{Balzer2013}.

\begin{figure}[t]
\centering
\includegraphics[width=0.7\textwidth]{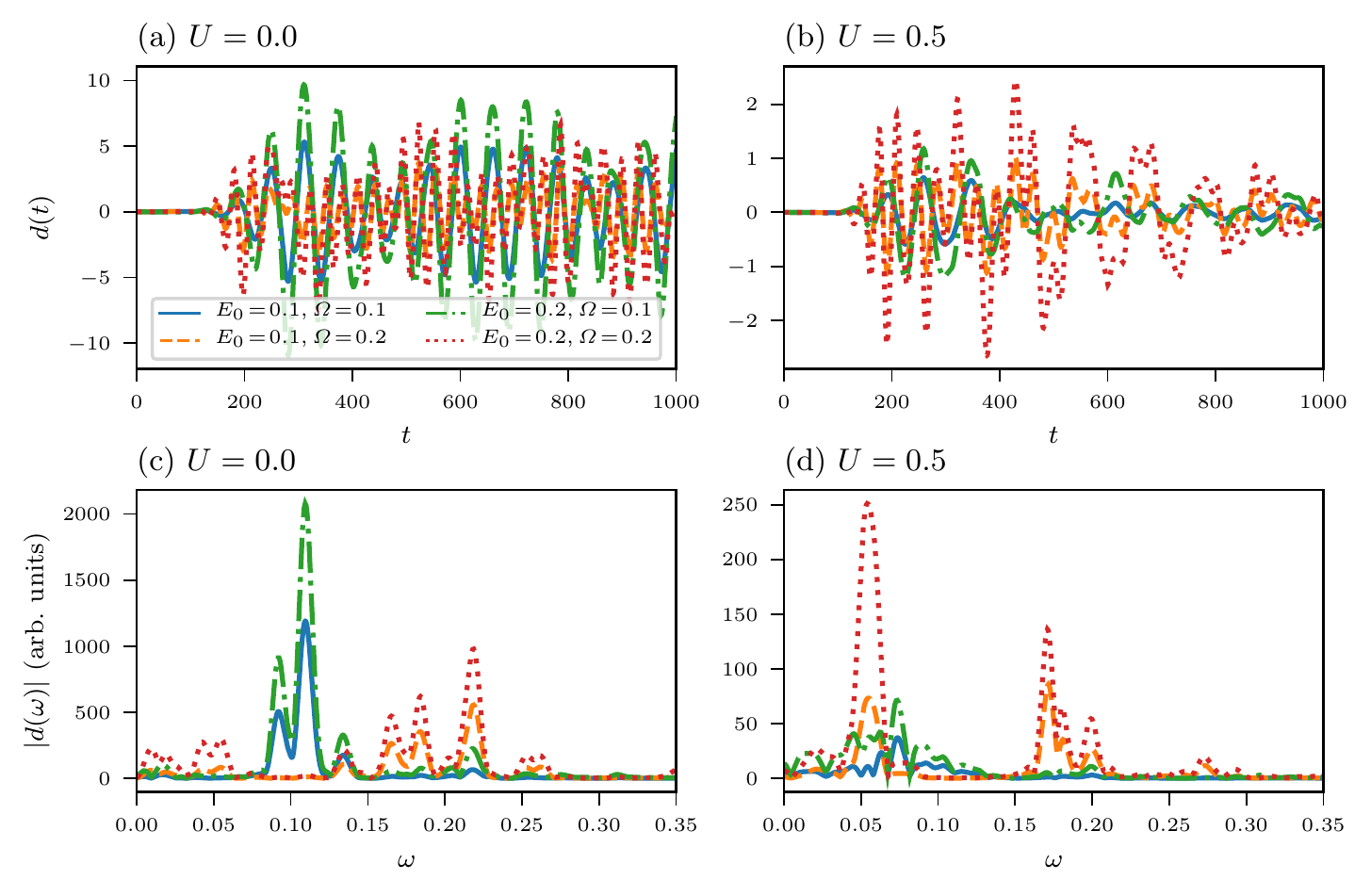}
\caption{Field-induced, time-dependent dipole moment for (a) noninteracting, $U=0.0$, and (b) interacting, $U=0.5$, nanowire, and the corresponding dipole spectrum (c-d) calculated as the absolute value of the Fourier transform of the time-dependent signals. The legend in panel (a) applies to all panels. The nanowire parameters are fixed $J=1$, $\a=0.5$, $\VZ=0.25$, $\varDelta=0.1$, $\mu=0$.}
\label{fig:pulse}
\end{figure}

Let us now investigate the time-dependent response of the nanowire in the topological regime ($J=1$, $\a=0.5$, $\VZ=0.25$, $\varDelta=0.1$, $\mu=0$) to a laser-pulse excitation. The field-induced, time-dependent dipole moment and the corresponding dipole spectra are shown in Figure~\ref{fig:pulse}. The laser-pulse excitation starts at $t=100$ and it is focused on the left-most atomic site of the nanowire [cf.~Equation~\eqref{eq:pulse}]. Irradiating different or larger portions of the nanowire has been checked to not alter the frequency content of the induced dynamics qualitatively. As the important mechanisms for the characterization of the MZM take place at low energies (in-gap states $\lesssim\varDelta$), we focus on low-frequency and low-amplitude pulses $\varOmega\in\{0.1,0.2\}$, $E_0\in\{0.1,0.2\}$ in Equation~\eqref{eq:pulse}. Applying such low-frequency pulses necessitates fairly long propagation times, for which the GKBA approach is beneficial. For a better frequency resolution, the Fourier transforms are calculated from an extended temporal window up to $t=1500$, and Blackman-window filtering is used~\cite{Blackman1959}.

The dipole moment oscillation amplitudes are considerably suppressed by the interactions, cf.~Figure~\ref{fig:pulse}(a-b). After the laser-pulse excitation, the charge redistribution along the nanowire is naturally affected by interparticle scatterings, and the back-and-forth charge sloshing is significantly damped. In the noninteracting case ($U=0.0$), the low-frequency pulse ($\varOmega=0.1$) excites only the first transitions from the zero-energy Majorana state to the first states around the superconducting gap edge, see Figure~\ref{fig:pulse}(c). This corresponds to the dominant oscillation seen in Figure~\ref{fig:pulse}(a). Some lower-frequency beating can also be observed although it is comparably weak. Higher-amplitude pulse only enhances the spectral peak heights but the peak locations remain. This picture is modified in the interacting case ($U=0.5$) as the double peak around $\w=0.1$ is spread over a range of smaller frequencies, see Figure~\ref{fig:pulse}(d). The main oscillatory character observed in Figure~\ref{fig:pulse}(a) is indeed suppressed in Figure~\ref{fig:pulse}(b), and there is no clear dominant transition as the interacting nanowire seems to contain more dipole-allowed transitions at lower energies. When the pulse frequency is increased to $\varOmega=0.2$, more transitions around the superconducting gap edge start to take place for the noninteracting case. These are visible as the low-frequency peaks in Figure~\ref{fig:pulse}(c). In addition, also higher-energy transitions are present. Interestingly, the interactions again modify this picture as the low-frequency peaks are bundled together around $\w=0.05$ and the higher-frequency ones between $\w\in[0.15, 0.20]$. In the interacting case, the higher-frequency driving now excites a few dominant transitions. This indicates a strong mixing of the spectral peaks of the interacting band structure (cf.~Figure~\ref{fig:kspec}) and the dipole matrix elements at these energies. While these excitations around the MZMs are accessible by low-frequency driving, it is likely that larger-frequency laser-pulse excitation would bring about high-order harmonics of the basic driving frequency and corresponding mixing with the dipole-allowed transitions~\cite{Myohanen2010,Ridley2017,Tuovinen2019nano}.

\subsection{Coupling to biased normal-metal leads}\label{sec:transport}

Ultimately, let us consider a quantum-transport setup with the nanowire being contacted to two normal-metal leads, cf.~Figure~\ref{fig:setup}. For this description, the Hamiltonian in Equation~\eqref{eq:hamiltonian} is supplemented with two additional terms for the leads and coupling~\cite{Tuovinen2020}
\begin{align}
\hat{H}_{\text{lead}} & = \sum_{k\lambda} \epsilon_{k\lambda} \hat{c}_{k\lambda}^\dagger \hat{c}_{k\lambda} , \\
\hat{H}_{\text{coupl}} & = \sum_{ik\lambda} (J_{ik\lambda} \hat{c}_i^\dagger \hat{c}_{k\lambda} + \hc) ,
\end{align}
respectively, where $\epsilon_{k\lambda}$ is the energy dispersion in lead $\lambda$ and $J_{ik\lambda}$ are the coupling matrix elements between the $i$-th site in the nanowire and the $k$-th basis function in lead $\lambda$. Again, like in Equation~\eqref{eq:hamiltonian}, the spin indices are suppressed and summed over. However, it would be possible to generalize this description to the spin-dependent case, such as for ferromagnetic leads. In accordance with the contour-time description in Section~\ref{sec:model}, the lead energy levels are shifted for times $t\geq t_0$ on the horizontal branch by $\epsilon_{k\lambda} \to \epsilon_{k\lambda} + V_\lambda(t)$ to model a bias-voltage profile. The leads are noninteracting which allows for a nonperturbative treatment via the embedding self-energy

\begin{align}
\varSigma_{\text{emb},\lambda}^{\text{R/A}}(t,t') & = \ex^{-\im \psi_\lambda(t,t')}\int\frac{\ud \w}{2\pi} \ex^{-\im\w (t-t')}[ \varLambda_\lambda(\w) \mp \im \varGamma_\lambda(\w)/2 ] , \label{eq:sigmaemret} \\
\varSigma_{\text{emb},\lambda}^{\lessgtr}(t,t') & = \pm \im \ex^{-\im \psi_\lambda(t,t')} \int \frac{\ud \w}{2 \pi} f[\pm(\w-\mu)] \varGamma_\lambda(\w) \ex^{-\im \w (t-t')} , \label{eq:sigmaemlss}
\end{align}
where $\psi_\lambda(t,t') = \int_{t'}^t \ud \bar{t} V_\lambda(\bar{t})$ is the bias-voltage phase factor and $f(x)=1/(\ex^{\beta x} + 1)$ is the Fermi function at inverse temperature $\beta$. The level-shift and level-width matrices are completely specified by the lead and coupling Hamiltonians
\begin{align}
(\varLambda_\lambda)_{ij}(\w) & = \sum_k J_{ik\lambda} \mathcal{P} \left(\frac{1}{\w-\epsilon_{k\lambda}}\right) J_{k\lambda j} , \label{eq:shift}\\
(\varGamma_\lambda)_{ij}(\w) & = 2\pi \sum_k J_{ik\lambda} \delta(\w-\epsilon_{k\lambda})J_{k\lambda j} , \label{eq:width}
\end{align}
where $\mathcal{P}$ denotes the principal value. In order to connect the retarded embedding self-energy with the retarded propagators in Equation~\eqref{eq:hf}, the wide-band limit approximation (WBLA) is considered. In this approximation, the level-width matrix is taken as frequency independent: $\varGamma_\lambda(\w) \approx \varGamma_\lambda$. Then, the level-shift matrix vanishes due to Kramers--Kronig relations, and the propagators are approximated by
\be\label{eq:hfemb}
G^{\text{R}/\text{A}}(t,t') \approx \mp \im \theta [\pm(t-t')]\mathcal{T}\ex^{-\im\int_{t'}^t \ud \bar{t}[h_{\text{HF}}(\bar{t})\mp \im\varGamma/2]},
\ee
where $\varGamma \equiv \sum_\lambda \varGamma_\lambda$. It is worth noticing, that the lesser/greater embedding self-energies in Equation~\eqref{eq:sigmaemlss} enter explicitly in the collision integral in Equation~\eqref{eq:collint} for which there is no requirement of WBLA. The WBLA is used only for the approximation of the propagators, and this approximation becomes better when Equations~\eqref{eq:shift} and~\eqref{eq:width} have weak dependence on frequency around the biased Fermi level of the leads.

Finally, after solving Equation~\eqref{eq:rho} with the addition of embedding self-energies in the collision integral and using Equation~\eqref{eq:gkba} together with Equation~\eqref{eq:hfemb}, the time-dependent current from lead $\lambda$ to the nanowire is calculated by the Meir--Wingreen formula~\cite{Meir1992}
\be\label{eq:mw}
I_\lambda(t) = 4 \mathrm{Re}\mathrm{Tr} \int_{t_0}^t \ud \bar{t} [ \varSigma_{\text{emb},\lambda}^>(t,\bar{t})G^<(\bar{t},t) - \varSigma_{\text{emb},\lambda}^<(t,\bar{t})G^>(\bar{t},t) ] .
\ee

\begin{figure}[t]
\centering
\includegraphics[width=0.7\textwidth]{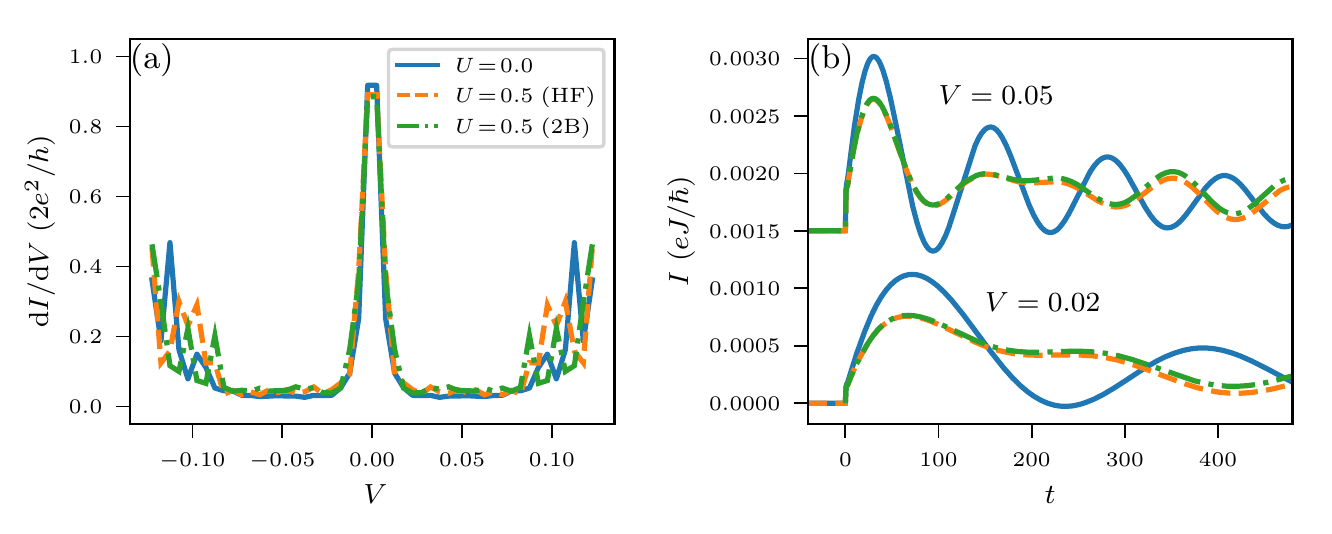}
\caption{Transport properties of the nanowire contacted to normal-metal leads. The tunneling rate is fixed such that $\varGamma_\lambda=0.01$ and the zero-temperature limit is considered. (a) Differential conductance obtained from the stationary current--voltage characteristics, where the bias voltage is applied symmetrically $V_L=-V_R \equiv V$. (b) Transient behavior of the current for two bias voltage values. For clarity, an upward shift of $0.0015$ is applied for the $V=0.05$ case. The legend in panel (a) applies to both panels. The nanowire parameters are fixed $J=1$, $\a=0.5$, $\VZ=0.25$, $\varDelta=0.1$, $\mu=0$.}
\label{fig:transport}
\end{figure}

As shown in Figure~\ref{fig:setup}, the nanowire is connected to left ($L$) and right ($R$) leads, i.e., $\lambda\in\{L,R\}$. The coupling strength from the first and $N$-th sites of the nanowire to the leads is chosen such that the tunneling rate $\varGamma_\lambda=0.01$. The bias-voltage profile is taken as a sudden shift of the lead energy levels, it is applied symmetrically, $V_L = -V_R \equiv V$, and the zero-temperature limit is considered. Stationary net current through the nanowire $I \equiv (I_L+I_R)/2$ is obtained from Equation~\eqref{eq:mw} at the long-time limit for various bias voltages. The bias-voltage window, $V \in [-0.125,0.125]$, is chosen relatively low for the description of the zero-energy Majorana states (in-gap states $\lesssim\varDelta$). In this regime, the WBLA is a very good approximation due to the weak coupling and small bias voltage~\cite{Zhu2005,Verzijl2013,Covito2018}. The stationary current is, in turn, used for the calculation of the differential conductance $\ud I/\ud V$ shown in Figure~\ref{fig:transport}(a). The differential conductance is peaked at the resonant energy levels of the nanowire within the bias-voltage window. In particular, the zero-bias peak associated with the MZM, is present for both noninteracting and interacting cases. This is consistent with the equilibrium spectral functions in Figure~\ref{fig:kspec}(a,c).

The current calculation also includes a comparison between the description of electronic interactions at the HF and 2B level. The interactions smear out the states around the gap edge so that transmission can be obtained with smaller bias voltages than in the noninteracting case. Qualitatively, HF and 2B give very similar results with each other; the zero-bias peak due to the MZM is also broadened similarly. In principle, this effect is not limited by the spectral broadening of the GKBA [cf.~Equation~\eqref{eq:hfemb}], because the current--voltage characteristics is obtained from the lesser/greater Green's function [cf.~Equation~\eqref{eq:mw}], which can contain more information than the retarded/advanced ones~\cite{Cosco2020}. However, as the interaction strength considered here is fairly small, no significant interaction-induced broadening of the differential conductance is observed, and it is mostly specified by the tunneling rate $\varGamma$.

It is worth pointing out that the long-time limit of Equation~\eqref{eq:mw} could be evaluated even without the proper relaxation of the initial state due to the loss of memory at the stationary state~\cite{Stefanucci2004}. However, for a proper description of the transient behavior, the preparation of the initial state is crucial. Not only the adiabatic switching of the many-body self-energy but also the embedding self-energy contributes to the relaxation of the initial state. For a relatively small tunneling rate, $\varGamma_\lambda=0.01$, it takes a fairly long relaxation time before the bias voltage can be applied. Recently, it has been shown that the information about the initial contact and correlation can be included in the out-of-equilibrium simulation as a separate calculation~\cite{Karlsson2018,Tuovinen2021}. This amounts to the inclusion of the imaginary-time convolution in Equation~\eqref{eq:collint} for both the correlation and embedding self-energies. In the present context, this is important both for the sake of efficient computation and for a partition-free treatment~\cite{Stefanucci2004, Ridley2018}.

The transient current through the nanowire is shown in Figure~\ref{fig:transport}(b) for two bias voltages $V \in \{0.02,0.05\}$ within the superconducting gap $\varDelta=0.1$. Therefore, only the in-gap MZMs are possible transport channels. The transient current grows rapidly and then starts oscillating. The relaxation towards the stationary current is relatively slow due to the weak coupling to the leads. In the noninteracting case, the oscillation frequencies correspond exactly to the energy difference between the biased Fermi level of the lead and the zero-energy state of the nanowire, cf.~Reference~\cite{Tuovinen2019NJP}. More precisely, for the bias voltage $V=0.05$ the oscillation period is roughly $125$ which translates to a frequency of $2\pi/125 \approx 0.05$. Accordingly, a slower oscillation is observed when $V=0.02$.

Electronic interactions, again, affect the transient behavior as the amplitude of the transient oscillations is damped due to the interparticle scattering events taking place throughout the nanowire. Also here, HF and 2B give qualitatively similar results with each other, and only the oscillation amplitudes are slightly altered. The absolute value of the current can therefore be altered by the approximation of the many-body self-energy but, clearly, it does not affect the differential conductance [cf.~Figure~\ref{fig:transport}(a)]. While the electronic interactions have already been seen to sustain the zero-energy Majorana state [cf.~Figure~\ref{fig:kspec}], the initial transient is affected by the interactions. After the initial transient has settled ($t\gtrsim 200$), the main oscillation, again, corresponds to the same transition from the biased Fermi level of the leads to the MZM in the nanowire. A proper Fourier analysis of the frequency spectrum, similar to Section~\ref{sec:pulse}, would require significantly longer time evolutions, which presently are still out of reach computationally. However, it can be generally argued that the transient oscillations of the current between the lead and the nanowire are qualitatively different than the overall charge (or plasma) oscillations within the nanowire, cf.~Section~\ref{sec:pulse}.

%%%%%%%%%%%%%%%%%%%%%%%%%%%%%%%%%%%%%%%%%%
\section{Conclusions}\label{sec:concl}

Electron-correlation effects in superconducting nanowires were studied in and out of equilibrium. The NEGF approach within the GKBA allowed for a simultaneous study of the correlation, embedding and transient effects. Particular emphasis was put on the role of electronic interactions in the topological superconducting phase and the associated MZMs.

In equilibrium, the MZMs were found to be protected against electronic interactions, and the equilibrium phase diagram to be extended to a larger chemical-potential window. This finding is in line with the DMRG data of Reference~\cite{Stoudenmire2011}, thus consolidating the applicability of the NEGF+GKBA approach for these systems.

Out of equilibrium, the transient build-up of the MZM was found to be affected by the electronic interactions. This was related to interparticle scattering events, taking place within the nanowire, which are obstructing the electronic signal. The transient charge oscillations excited by a laser pulse were also found to be damped due to interactions. On the other hand, time-resolved transport signatures were found to be qualitatively less affected by the interactions because, in that case, the main transient oscillations resulted from transitions between the biased Fermi level of the leads and the zero-energy states within the nanowire~\cite{Tuovinen2019NJP}. In general, the transient oscillations carry important information about the underlying out-of-equilibrium scattering mechanisms, which might not be available from the stationary-state data.

While the experimental verification of the MZMs in these systems is yet to be presented, it is useful to estimate limits for the required temporal resolution for time-resolved transport measurements with the help of the simulations presented here. As the unit system was fixed by the hopping value $J=1$, which determines, e.g., the material's bandwidth and is typically on the electron-volt scale, the transient current oscillations lasting for hundreds of $J^{-1}$ would correspond to the picosecond time scale. While this resolution is at limits of what is routinely achievable, recent development in ultrafast transport measurements with on-chip femtosecond technology has allowed a sub-picosecond temporal resolution to detect the Hall current in graphene~\cite{McIver2020}.

The computational effort for the resolution of the out-of-equilibrium simulations is not to be underestimated. Due to the relatively large spin$\otimes$particle--hole basis of the system studied here, the efficient construction of the many-body self-energies is important~\cite{Tuovinen2019JCP}. While the time-propagation via the GKBA approach considered here scales as the number of time steps squared (compared to the cubic scaling of the full Kadanoff--Baym equations), it may still render longer time evolutions fairly inaccessible. Recent progress in this issue has allowed for an equivalent but more efficient representation of the GKBA time evolution with only a linear scaling in the number of time steps~\cite{Schluenzen2020PRL, Joost2020, Karlsson2020, Pavlyukh2021}. It would be very useful to extend these procedures to open quantum systems in order to study, e.g., time-dependent radiation in molecular junctions out-of-equilibrium~\cite{Zhang2020,Ridley2021}. This shall be addressed in a forthcoming paper.

\acknowledgments

This research was funded by the Academy of Finland Project No. 345007.
CSC--IT Center for Science, Finland, is acknowledged for computational resources.

\section*{Data availability}
The data that support the findings of this study are available from the author upon reasonable request.


\begin{thebibliography}{79}%
\makeatletter
\providecommand \@ifxundefined [1]{%
 \@ifx{#1\undefined}
}%
\providecommand \@ifnum [1]{%
 \ifnum #1\expandafter \@firstoftwo
 \else \expandafter \@secondoftwo
 \fi
}%
\providecommand \@ifx [1]{%
 \ifx #1\expandafter \@firstoftwo
 \else \expandafter \@secondoftwo
 \fi
}%
\providecommand \natexlab [1]{#1}%
\providecommand \enquote  [1]{``#1''}%
\providecommand \bibnamefont  [1]{#1}%
\providecommand \bibfnamefont [1]{#1}%
\providecommand \citenamefont [1]{#1}%
\providecommand \href@noop [0]{\@secondoftwo}%
\providecommand \href [0]{\begingroup \@sanitize@url \@href}%
\providecommand \@href[1]{\@@startlink{#1}\@@href}%
\providecommand \@@href[1]{\endgroup#1\@@endlink}%
\providecommand \@sanitize@url [0]{\catcode `\\12\catcode `\$12\catcode
  `\&12\catcode `\#12\catcode `\^12\catcode `\_12\catcode `\%12\relax}%
\providecommand \@@startlink[1]{}%
\providecommand \@@endlink[0]{}%
\providecommand \url  [0]{\begingroup\@sanitize@url \@url }%
\providecommand \@url [1]{\endgroup\@href {#1}{\urlprefix }}%
\providecommand \urlprefix  [0]{URL }%
\providecommand \Eprint [0]{\href }%
\providecommand \doibase [0]{http://dx.doi.org/}%
\providecommand \selectlanguage [0]{\@gobble}%
\providecommand \bibinfo  [0]{\@secondoftwo}%
\providecommand \bibfield  [0]{\@secondoftwo}%
\providecommand \translation [1]{[#1]}%
\providecommand \BibitemOpen [0]{}%
\providecommand \bibitemStop [0]{}%
\providecommand \bibitemNoStop [0]{.\EOS\space}%
\providecommand \EOS [0]{\spacefactor3000\relax}%
\providecommand \BibitemShut  [1]{\csname bibitem#1\endcsname}%
\let\auto@bib@innerbib\@empty
%</preamble>
\bibitem [{\citenamefont {Oreg}\ \emph {et~al.}(2010)\citenamefont {Oreg},
  \citenamefont {Refael},\ and\ \citenamefont {von Oppen}}]{Oreg2010}%
  \BibitemOpen
  \bibfield  {author} {\bibinfo {author} {\bibfnamefont {Y.}~\bibnamefont
  {Oreg}}, \bibinfo {author} {\bibfnamefont {G.}~\bibnamefont {Refael}}, \ and\
  \bibinfo {author} {\bibfnamefont {F.}~\bibnamefont {von Oppen}},\ }\href
  {\doibase 10.1103/PhysRevLett.105.177002} {\bibfield  {journal} {\bibinfo
  {journal} {Phys. Rev. Lett.}\ }\textbf {\bibinfo {volume} {105}},\ \bibinfo
  {pages} {177002} (\bibinfo {year} {2010})}\BibitemShut {NoStop}%
\bibitem [{\citenamefont {Lutchyn}\ \emph {et~al.}(2010)\citenamefont
  {Lutchyn}, \citenamefont {Sau},\ and\ \citenamefont
  {Das~Sarma}}]{Lutchyn2010}%
  \BibitemOpen
  \bibfield  {author} {\bibinfo {author} {\bibfnamefont {R.~M.}\ \bibnamefont
  {Lutchyn}}, \bibinfo {author} {\bibfnamefont {J.~D.}\ \bibnamefont {Sau}}, \
  and\ \bibinfo {author} {\bibfnamefont {S.}~\bibnamefont {Das~Sarma}},\ }\href
  {\doibase 10.1103/PhysRevLett.105.077001} {\bibfield  {journal} {\bibinfo
  {journal} {Phys. Rev. Lett.}\ }\textbf {\bibinfo {volume} {105}},\ \bibinfo
  {pages} {077001} (\bibinfo {year} {2010})}\BibitemShut {NoStop}%
\bibitem [{\citenamefont {Kitaev}(2003)}]{Kitaev2003}%
  \BibitemOpen
  \bibfield  {author} {\bibinfo {author} {\bibfnamefont {A.}~\bibnamefont
  {Kitaev}},\ }\href {\doibase 10.1016/S0003-4916(02)00018-0} {\bibfield
  {journal} {\bibinfo  {journal} {Ann. Phys.}\ }\textbf {\bibinfo {volume}
  {303}},\ \bibinfo {pages} {2} (\bibinfo {year} {2003})}\BibitemShut {NoStop}%
\bibitem [{\citenamefont {Nayak}\ \emph {et~al.}(2008)\citenamefont {Nayak},
  \citenamefont {Simon}, \citenamefont {Stern}, \citenamefont {Freedman},\ and\
  \citenamefont {Das~Sarma}}]{Nayak2008}%
  \BibitemOpen
  \bibfield  {author} {\bibinfo {author} {\bibfnamefont {C.}~\bibnamefont
  {Nayak}}, \bibinfo {author} {\bibfnamefont {S.~H.}\ \bibnamefont {Simon}},
  \bibinfo {author} {\bibfnamefont {A.}~\bibnamefont {Stern}}, \bibinfo
  {author} {\bibfnamefont {M.}~\bibnamefont {Freedman}}, \ and\ \bibinfo
  {author} {\bibfnamefont {S.}~\bibnamefont {Das~Sarma}},\ }\href {\doibase
  10.1103/RevModPhys.80.1083} {\bibfield  {journal} {\bibinfo  {journal} {Rev.
  Mod. Phys.}\ }\textbf {\bibinfo {volume} {80}},\ \bibinfo {pages} {1083}
  (\bibinfo {year} {2008})}\BibitemShut {NoStop}%
\bibitem [{\citenamefont {Mourik}\ \emph {et~al.}(2012)\citenamefont {Mourik},
  \citenamefont {Zuo}, \citenamefont {Frolov}, \citenamefont {Plissard},
  \citenamefont {Bakkers},\ and\ \citenamefont {Kouwenhoven}}]{Mourik2012}%
  \BibitemOpen
  \bibfield  {author} {\bibinfo {author} {\bibfnamefont {V.}~\bibnamefont
  {Mourik}}, \bibinfo {author} {\bibfnamefont {K.}~\bibnamefont {Zuo}},
  \bibinfo {author} {\bibfnamefont {S.~M.}\ \bibnamefont {Frolov}}, \bibinfo
  {author} {\bibfnamefont {S.~R.}\ \bibnamefont {Plissard}}, \bibinfo {author}
  {\bibfnamefont {E.~P. A.~M.}\ \bibnamefont {Bakkers}}, \ and\ \bibinfo
  {author} {\bibfnamefont {L.~P.}\ \bibnamefont {Kouwenhoven}},\ }\href
  {\doibase 10.1126/science.1222360} {\bibfield  {journal} {\bibinfo  {journal}
  {Science}\ }\textbf {\bibinfo {volume} {336}},\ \bibinfo {pages} {1003}
  (\bibinfo {year} {2012})}\BibitemShut {NoStop}%
\bibitem [{\citenamefont {Suominen}\ \emph {et~al.}(2017)\citenamefont
  {Suominen}, \citenamefont {Kjaergaard}, \citenamefont {Hamilton},
  \citenamefont {Shabani}, \citenamefont {Palmstr\o{}m}, \citenamefont
  {Marcus},\ and\ \citenamefont {Nichele}}]{Suominen2017}%
  \BibitemOpen
  \bibfield  {author} {\bibinfo {author} {\bibfnamefont {H.~J.}\ \bibnamefont
  {Suominen}}, \bibinfo {author} {\bibfnamefont {M.}~\bibnamefont
  {Kjaergaard}}, \bibinfo {author} {\bibfnamefont {A.~R.}\ \bibnamefont
  {Hamilton}}, \bibinfo {author} {\bibfnamefont {J.}~\bibnamefont {Shabani}},
  \bibinfo {author} {\bibfnamefont {C.~J.}\ \bibnamefont {Palmstr\o{}m}},
  \bibinfo {author} {\bibfnamefont {C.~M.}\ \bibnamefont {Marcus}}, \ and\
  \bibinfo {author} {\bibfnamefont {F.}~\bibnamefont {Nichele}},\ }\href
  {\doibase 10.1103/PhysRevLett.119.176805} {\bibfield  {journal} {\bibinfo
  {journal} {Phys. Rev. Lett.}\ }\textbf {\bibinfo {volume} {119}},\ \bibinfo
  {pages} {176805} (\bibinfo {year} {2017})}\BibitemShut {NoStop}%
\bibitem [{\citenamefont {Prada}\ \emph {et~al.}(2020)\citenamefont {Prada},
  \citenamefont {San-Jose}, \citenamefont {de~Moor}, \citenamefont {Geresdi},
  \citenamefont {Lee}, \citenamefont {Klinovaja}, \citenamefont {Loss},
  \citenamefont {Nyg{\aa}rd}, \citenamefont {Aguado},\ and\ \citenamefont
  {Kouwenhoven}}]{Prada2020}%
  \BibitemOpen
  \bibfield  {author} {\bibinfo {author} {\bibfnamefont {E.}~\bibnamefont
  {Prada}}, \bibinfo {author} {\bibfnamefont {P.}~\bibnamefont {San-Jose}},
  \bibinfo {author} {\bibfnamefont {M.~W.~A.}\ \bibnamefont {de~Moor}},
  \bibinfo {author} {\bibfnamefont {A.}~\bibnamefont {Geresdi}}, \bibinfo
  {author} {\bibfnamefont {E.~J.~H.}\ \bibnamefont {Lee}}, \bibinfo {author}
  {\bibfnamefont {J.}~\bibnamefont {Klinovaja}}, \bibinfo {author}
  {\bibfnamefont {D.}~\bibnamefont {Loss}}, \bibinfo {author} {\bibfnamefont
  {J.}~\bibnamefont {Nyg{\aa}rd}}, \bibinfo {author} {\bibfnamefont
  {R.}~\bibnamefont {Aguado}}, \ and\ \bibinfo {author} {\bibfnamefont {L.~P.}\
  \bibnamefont {Kouwenhoven}},\ }\href {\doibase 10.1038/s42254-020-0228-y}
  {\bibfield  {journal} {\bibinfo  {journal} {Nat. Rev. Phys.}\ }\textbf
  {\bibinfo {volume} {2}},\ \bibinfo {pages} {575} (\bibinfo {year}
  {2020})}\BibitemShut {NoStop}%
\bibitem [{\citenamefont {Yu}\ \emph {et~al.}(2021)\citenamefont {Yu},
  \citenamefont {Chen}, \citenamefont {Gomanko}, \citenamefont {Badawy},
  \citenamefont {Bakkers}, \citenamefont {Zuo}, \citenamefont {Mourik},\ and\
  \citenamefont {Frolov}}]{Yu2021}%
  \BibitemOpen
  \bibfield  {author} {\bibinfo {author} {\bibfnamefont {P.}~\bibnamefont
  {Yu}}, \bibinfo {author} {\bibfnamefont {J.}~\bibnamefont {Chen}}, \bibinfo
  {author} {\bibfnamefont {M.}~\bibnamefont {Gomanko}}, \bibinfo {author}
  {\bibfnamefont {G.}~\bibnamefont {Badawy}}, \bibinfo {author} {\bibfnamefont
  {E.~P. A.~M.}\ \bibnamefont {Bakkers}}, \bibinfo {author} {\bibfnamefont
  {K.}~\bibnamefont {Zuo}}, \bibinfo {author} {\bibfnamefont {V.}~\bibnamefont
  {Mourik}}, \ and\ \bibinfo {author} {\bibfnamefont {S.~M.}\ \bibnamefont
  {Frolov}},\ }\href {\doibase 10.1038/s41567-020-01107-w} {\bibfield
  {journal} {\bibinfo  {journal} {Nat. Phys.}\ }\textbf {\bibinfo {volume}
  {17}},\ \bibinfo {pages} {482} (\bibinfo {year} {2021})}\BibitemShut
  {NoStop}%
\bibitem [{\citenamefont {Lobos}\ \emph {et~al.}(2012)\citenamefont {Lobos},
  \citenamefont {Lutchyn},\ and\ \citenamefont {Das~Sarma}}]{Lobos2012}%
  \BibitemOpen
  \bibfield  {author} {\bibinfo {author} {\bibfnamefont {A.~M.}\ \bibnamefont
  {Lobos}}, \bibinfo {author} {\bibfnamefont {R.~M.}\ \bibnamefont {Lutchyn}},
  \ and\ \bibinfo {author} {\bibfnamefont {S.}~\bibnamefont {Das~Sarma}},\
  }\href {\doibase 10.1103/PhysRevLett.109.146403} {\bibfield  {journal}
  {\bibinfo  {journal} {Phys. Rev. Lett.}\ }\textbf {\bibinfo {volume} {109}},\
  \bibinfo {pages} {146403} (\bibinfo {year} {2012})}\BibitemShut {NoStop}%
\bibitem [{\citenamefont {Sau}\ and\ \citenamefont
  {Das~Sarma}(2013)}]{Sau2013}%
  \BibitemOpen
  \bibfield  {author} {\bibinfo {author} {\bibfnamefont {J.~D.}\ \bibnamefont
  {Sau}}\ and\ \bibinfo {author} {\bibfnamefont {S.}~\bibnamefont
  {Das~Sarma}},\ }\href {\doibase 10.1103/PhysRevB.88.064506} {\bibfield
  {journal} {\bibinfo  {journal} {Phys. Rev. B}\ }\textbf {\bibinfo {volume}
  {88}},\ \bibinfo {pages} {064506} (\bibinfo {year} {2013})}\BibitemShut
  {NoStop}%
\bibitem [{\citenamefont {Cole}\ \emph {et~al.}(2016)\citenamefont {Cole},
  \citenamefont {Sau},\ and\ \citenamefont {Das~Sarma}}]{Cole2016}%
  \BibitemOpen
  \bibfield  {author} {\bibinfo {author} {\bibfnamefont {W.~S.}\ \bibnamefont
  {Cole}}, \bibinfo {author} {\bibfnamefont {J.~D.}\ \bibnamefont {Sau}}, \
  and\ \bibinfo {author} {\bibfnamefont {S.}~\bibnamefont {Das~Sarma}},\ }\href
  {\doibase 10.1103/PhysRevB.94.140505} {\bibfield  {journal} {\bibinfo
  {journal} {Phys. Rev. B}\ }\textbf {\bibinfo {volume} {94}},\ \bibinfo
  {pages} {140505} (\bibinfo {year} {2016})}\BibitemShut {NoStop}%
\bibitem [{\citenamefont {Kaladzhyan}\ \emph {et~al.}(2016)\citenamefont
  {Kaladzhyan}, \citenamefont {R\"ontynen}, \citenamefont {Simon},\ and\
  \citenamefont {Ojanen}}]{Kaladzhyan2016}%
  \BibitemOpen
  \bibfield  {author} {\bibinfo {author} {\bibfnamefont {V.}~\bibnamefont
  {Kaladzhyan}}, \bibinfo {author} {\bibfnamefont {J.}~\bibnamefont
  {R\"ontynen}}, \bibinfo {author} {\bibfnamefont {P.}~\bibnamefont {Simon}}, \
  and\ \bibinfo {author} {\bibfnamefont {T.}~\bibnamefont {Ojanen}},\ }\href
  {\doibase 10.1103/PhysRevB.94.060505} {\bibfield  {journal} {\bibinfo
  {journal} {Phys. Rev. B}\ }\textbf {\bibinfo {volume} {94}},\ \bibinfo
  {pages} {060505} (\bibinfo {year} {2016})}\BibitemShut {NoStop}%
\bibitem [{\citenamefont {Andolina}\ and\ \citenamefont
  {Simon}(2017)}]{Andolina2017}%
  \BibitemOpen
  \bibfield  {author} {\bibinfo {author} {\bibfnamefont {G.~M.}\ \bibnamefont
  {Andolina}}\ and\ \bibinfo {author} {\bibfnamefont {P.}~\bibnamefont
  {Simon}},\ }\href {\doibase 10.1103/PhysRevB.96.235411} {\bibfield  {journal}
  {\bibinfo  {journal} {Phys. Rev. B}\ }\textbf {\bibinfo {volume} {96}},\
  \bibinfo {pages} {235411} (\bibinfo {year} {2017})}\BibitemShut {NoStop}%
\bibitem [{\citenamefont {Antipov}\ \emph {et~al.}(2018)\citenamefont
  {Antipov}, \citenamefont {Bargerbos}, \citenamefont {Winkler}, \citenamefont
  {Bauer}, \citenamefont {Rossi},\ and\ \citenamefont {Lutchyn}}]{Antipov2018}%
  \BibitemOpen
  \bibfield  {author} {\bibinfo {author} {\bibfnamefont {A.~E.}\ \bibnamefont
  {Antipov}}, \bibinfo {author} {\bibfnamefont {A.}~\bibnamefont {Bargerbos}},
  \bibinfo {author} {\bibfnamefont {G.~W.}\ \bibnamefont {Winkler}}, \bibinfo
  {author} {\bibfnamefont {B.}~\bibnamefont {Bauer}}, \bibinfo {author}
  {\bibfnamefont {E.}~\bibnamefont {Rossi}}, \ and\ \bibinfo {author}
  {\bibfnamefont {R.~M.}\ \bibnamefont {Lutchyn}},\ }\href {\doibase
  10.1103/PhysRevX.8.031041} {\bibfield  {journal} {\bibinfo  {journal} {Phys.
  Rev. X}\ }\textbf {\bibinfo {volume} {8}},\ \bibinfo {pages} {031041}
  (\bibinfo {year} {2018})}\BibitemShut {NoStop}%
\bibitem [{\citenamefont {Liu}\ \emph {et~al.}(2018)\citenamefont {Liu},
  \citenamefont {Rossi},\ and\ \citenamefont {Lutchyn}}]{Liu2018}%
  \BibitemOpen
  \bibfield  {author} {\bibinfo {author} {\bibfnamefont {D.~E.}\ \bibnamefont
  {Liu}}, \bibinfo {author} {\bibfnamefont {E.}~\bibnamefont {Rossi}}, \ and\
  \bibinfo {author} {\bibfnamefont {R.~M.}\ \bibnamefont {Lutchyn}},\ }\href
  {\doibase 10.1103/PhysRevB.97.161408} {\bibfield  {journal} {\bibinfo
  {journal} {Phys. Rev. B}\ }\textbf {\bibinfo {volume} {97}},\ \bibinfo
  {pages} {161408} (\bibinfo {year} {2018})}\BibitemShut {NoStop}%
\bibitem [{\citenamefont {Thakurathi}\ \emph {et~al.}(2018)\citenamefont
  {Thakurathi}, \citenamefont {Simon}, \citenamefont {Mandal}, \citenamefont
  {Klinovaja},\ and\ \citenamefont {Loss}}]{Thakurathi2018}%
  \BibitemOpen
  \bibfield  {author} {\bibinfo {author} {\bibfnamefont {M.}~\bibnamefont
  {Thakurathi}}, \bibinfo {author} {\bibfnamefont {P.}~\bibnamefont {Simon}},
  \bibinfo {author} {\bibfnamefont {I.}~\bibnamefont {Mandal}}, \bibinfo
  {author} {\bibfnamefont {J.}~\bibnamefont {Klinovaja}}, \ and\ \bibinfo
  {author} {\bibfnamefont {D.}~\bibnamefont {Loss}},\ }\href {\doibase
  10.1103/PhysRevB.97.045415} {\bibfield  {journal} {\bibinfo  {journal} {Phys.
  Rev. B}\ }\textbf {\bibinfo {volume} {97}},\ \bibinfo {pages} {045415}
  (\bibinfo {year} {2018})}\BibitemShut {NoStop}%
\bibitem [{\citenamefont {Pan}\ and\ \citenamefont
  {Das~Sarma}(2020)}]{Pan2020}%
  \BibitemOpen
  \bibfield  {author} {\bibinfo {author} {\bibfnamefont {H.}~\bibnamefont
  {Pan}}\ and\ \bibinfo {author} {\bibfnamefont {S.}~\bibnamefont
  {Das~Sarma}},\ }\href {\doibase 10.1103/PhysRevResearch.2.013377} {\bibfield
  {journal} {\bibinfo  {journal} {Phys. Rev. Research}\ }\textbf {\bibinfo
  {volume} {2}},\ \bibinfo {pages} {013377} (\bibinfo {year}
  {2020})}\BibitemShut {NoStop}%
\bibitem [{\citenamefont {Gangadharaiah}\ \emph {et~al.}(2011)\citenamefont
  {Gangadharaiah}, \citenamefont {Braunecker}, \citenamefont {Simon},\ and\
  \citenamefont {Loss}}]{Gangadharaiah2011}%
  \BibitemOpen
  \bibfield  {author} {\bibinfo {author} {\bibfnamefont {S.}~\bibnamefont
  {Gangadharaiah}}, \bibinfo {author} {\bibfnamefont {B.}~\bibnamefont
  {Braunecker}}, \bibinfo {author} {\bibfnamefont {P.}~\bibnamefont {Simon}}, \
  and\ \bibinfo {author} {\bibfnamefont {D.}~\bibnamefont {Loss}},\ }\href
  {\doibase 10.1103/PhysRevLett.107.036801} {\bibfield  {journal} {\bibinfo
  {journal} {Phys. Rev. Lett.}\ }\textbf {\bibinfo {volume} {107}},\ \bibinfo
  {pages} {036801} (\bibinfo {year} {2011})}\BibitemShut {NoStop}%
\bibitem [{\citenamefont {Stoudenmire}\ \emph {et~al.}(2011)\citenamefont
  {Stoudenmire}, \citenamefont {Alicea}, \citenamefont {Starykh},\ and\
  \citenamefont {Fisher}}]{Stoudenmire2011}%
  \BibitemOpen
  \bibfield  {author} {\bibinfo {author} {\bibfnamefont {E.~M.}\ \bibnamefont
  {Stoudenmire}}, \bibinfo {author} {\bibfnamefont {J.}~\bibnamefont {Alicea}},
  \bibinfo {author} {\bibfnamefont {O.~A.}\ \bibnamefont {Starykh}}, \ and\
  \bibinfo {author} {\bibfnamefont {M.~P.}\ \bibnamefont {Fisher}},\ }\href
  {\doibase 10.1103/PhysRevB.84.014503} {\bibfield  {journal} {\bibinfo
  {journal} {Phys. Rev. B}\ }\textbf {\bibinfo {volume} {84}},\ \bibinfo
  {pages} {014503} (\bibinfo {year} {2011})}\BibitemShut {NoStop}%
\bibitem [{\citenamefont {Sela}\ \emph {et~al.}(2011)\citenamefont {Sela},
  \citenamefont {Altland},\ and\ \citenamefont {Rosch}}]{Sela2011}%
  \BibitemOpen
  \bibfield  {author} {\bibinfo {author} {\bibfnamefont {E.}~\bibnamefont
  {Sela}}, \bibinfo {author} {\bibfnamefont {A.}~\bibnamefont {Altland}}, \
  and\ \bibinfo {author} {\bibfnamefont {A.}~\bibnamefont {Rosch}},\ }\href
  {\doibase 10.1103/PhysRevB.84.085114} {\bibfield  {journal} {\bibinfo
  {journal} {Phys. Rev. B}\ }\textbf {\bibinfo {volume} {84}},\ \bibinfo
  {pages} {085114} (\bibinfo {year} {2011})}\BibitemShut {NoStop}%
\bibitem [{\citenamefont {Haim}\ \emph {et~al.}(2014)\citenamefont {Haim},
  \citenamefont {Keselman}, \citenamefont {Berg},\ and\ \citenamefont
  {Oreg}}]{Haim2014}%
  \BibitemOpen
  \bibfield  {author} {\bibinfo {author} {\bibfnamefont {A.}~\bibnamefont
  {Haim}}, \bibinfo {author} {\bibfnamefont {A.}~\bibnamefont {Keselman}},
  \bibinfo {author} {\bibfnamefont {E.}~\bibnamefont {Berg}}, \ and\ \bibinfo
  {author} {\bibfnamefont {Y.}~\bibnamefont {Oreg}},\ }\href {\doibase
  10.1103/PhysRevB.89.220504} {\bibfield  {journal} {\bibinfo  {journal} {Phys.
  Rev. B}\ }\textbf {\bibinfo {volume} {89}},\ \bibinfo {pages} {220504}
  (\bibinfo {year} {2014})}\BibitemShut {NoStop}%
\bibitem [{\citenamefont {Sticlet}\ \emph {et~al.}(2014)\citenamefont
  {Sticlet}, \citenamefont {Seabra}, \citenamefont {Pollmann},\ and\
  \citenamefont {Cayssol}}]{Sticlet2014}%
  \BibitemOpen
  \bibfield  {author} {\bibinfo {author} {\bibfnamefont {D.}~\bibnamefont
  {Sticlet}}, \bibinfo {author} {\bibfnamefont {L.}~\bibnamefont {Seabra}},
  \bibinfo {author} {\bibfnamefont {F.}~\bibnamefont {Pollmann}}, \ and\
  \bibinfo {author} {\bibfnamefont {J.}~\bibnamefont {Cayssol}},\ }\href
  {\doibase 10.1103/PhysRevB.89.115430} {\bibfield  {journal} {\bibinfo
  {journal} {Phys. Rev. B}\ }\textbf {\bibinfo {volume} {89}},\ \bibinfo
  {pages} {115430} (\bibinfo {year} {2014})}\BibitemShut {NoStop}%
\bibitem [{\citenamefont {Ruiz-Tijerina}\ \emph {et~al.}(2015)\citenamefont
  {Ruiz-Tijerina}, \citenamefont {Vernek}, \citenamefont {Dias~da Silva},\ and\
  \citenamefont {Egues}}]{Ruiz2015}%
  \BibitemOpen
  \bibfield  {author} {\bibinfo {author} {\bibfnamefont {D.~A.}\ \bibnamefont
  {Ruiz-Tijerina}}, \bibinfo {author} {\bibfnamefont {E.}~\bibnamefont
  {Vernek}}, \bibinfo {author} {\bibfnamefont {L.~G. G.~V.}\ \bibnamefont
  {Dias~da Silva}}, \ and\ \bibinfo {author} {\bibfnamefont {J.~C.}\
  \bibnamefont {Egues}},\ }\href {\doibase 10.1103/PhysRevB.91.115435}
  {\bibfield  {journal} {\bibinfo  {journal} {Phys. Rev. B}\ }\textbf {\bibinfo
  {volume} {91}},\ \bibinfo {pages} {115435} (\bibinfo {year}
  {2015})}\BibitemShut {NoStop}%
\bibitem [{\citenamefont {Haim}\ \emph {et~al.}(2016)\citenamefont {Haim},
  \citenamefont {W\"olms}, \citenamefont {Berg}, \citenamefont {Oreg},\ and\
  \citenamefont {Flensberg}}]{Haim2016}%
  \BibitemOpen
  \bibfield  {author} {\bibinfo {author} {\bibfnamefont {A.}~\bibnamefont
  {Haim}}, \bibinfo {author} {\bibfnamefont {K.}~\bibnamefont {W\"olms}},
  \bibinfo {author} {\bibfnamefont {E.}~\bibnamefont {Berg}}, \bibinfo {author}
  {\bibfnamefont {Y.}~\bibnamefont {Oreg}}, \ and\ \bibinfo {author}
  {\bibfnamefont {K.}~\bibnamefont {Flensberg}},\ }\href {\doibase
  10.1103/PhysRevB.94.115124} {\bibfield  {journal} {\bibinfo  {journal} {Phys.
  Rev. B}\ }\textbf {\bibinfo {volume} {94}},\ \bibinfo {pages} {115124}
  (\bibinfo {year} {2016})}\BibitemShut {NoStop}%
\bibitem [{\citenamefont {Dom{\'i}nguez}\ \emph {et~al.}(2017)\citenamefont
  {Dom{\'i}nguez}, \citenamefont {Cayao}, \citenamefont {San-Jose},
  \citenamefont {Aguado}, \citenamefont {Yeyati},\ and\ \citenamefont
  {Prada}}]{Dominguez2017}%
  \BibitemOpen
  \bibfield  {author} {\bibinfo {author} {\bibfnamefont {F.}~\bibnamefont
  {Dom{\'i}nguez}}, \bibinfo {author} {\bibfnamefont {J.}~\bibnamefont
  {Cayao}}, \bibinfo {author} {\bibfnamefont {P.}~\bibnamefont {San-Jose}},
  \bibinfo {author} {\bibfnamefont {R.}~\bibnamefont {Aguado}}, \bibinfo
  {author} {\bibfnamefont {A.~L.}\ \bibnamefont {Yeyati}}, \ and\ \bibinfo
  {author} {\bibfnamefont {E.}~\bibnamefont {Prada}},\ }\href {\doibase
  10.1038/s41535-017-0012-0} {\bibfield  {journal} {\bibinfo  {journal} {npj
  Quantum Mater.}\ }\textbf {\bibinfo {volume} {2}},\ \bibinfo {pages} {13}
  (\bibinfo {year} {2017})}\BibitemShut {NoStop}%
\bibitem [{\citenamefont {Li}\ \emph {et~al.}(2019)\citenamefont {Li},
  \citenamefont {Burrello},\ and\ \citenamefont {Flensberg}}]{Li2019}%
  \BibitemOpen
  \bibfield  {author} {\bibinfo {author} {\bibfnamefont {T.}~\bibnamefont
  {Li}}, \bibinfo {author} {\bibfnamefont {M.}~\bibnamefont {Burrello}}, \ and\
  \bibinfo {author} {\bibfnamefont {K.}~\bibnamefont {Flensberg}},\ }\href
  {\doibase 10.1103/PhysRevB.100.045305} {\bibfield  {journal} {\bibinfo
  {journal} {Phys. Rev. B}\ }\textbf {\bibinfo {volume} {100}},\ \bibinfo
  {pages} {045305} (\bibinfo {year} {2019})}\BibitemShut {NoStop}%
\bibitem [{\citenamefont {Wang}\ \emph {et~al.}(2020)\citenamefont {Wang},
  \citenamefont {Wu}, \citenamefont {Zhang}, \citenamefont {Wang},\ and\
  \citenamefont {Gong}}]{Wang2020}%
  \BibitemOpen
  \bibfield  {author} {\bibinfo {author} {\bibfnamefont {X.-Q.}\ \bibnamefont
  {Wang}}, \bibinfo {author} {\bibfnamefont {B.}~\bibnamefont {Wu}}, \bibinfo
  {author} {\bibfnamefont {S.-F.}\ \bibnamefont {Zhang}}, \bibinfo {author}
  {\bibfnamefont {Q.}~\bibnamefont {Wang}}, \ and\ \bibinfo {author}
  {\bibfnamefont {W.-J.}\ \bibnamefont {Gong}},\ }\href {\doibase
  10.1016/j.aop.2020.168127} {\bibfield  {journal} {\bibinfo  {journal} {Ann.
  Phys.}\ }\textbf {\bibinfo {volume} {415}},\ \bibinfo {pages} {168127}
  (\bibinfo {year} {2020})}\BibitemShut {NoStop}%
\bibitem [{\citenamefont {Vadimov}\ \emph {et~al.}(2021)\citenamefont
  {Vadimov}, \citenamefont {Hyart}, \citenamefont {Lado}, \citenamefont
  {M\"ott\"onen},\ and\ \citenamefont {Ala-Nissila}}]{Vadimov2021}%
  \BibitemOpen
  \bibfield  {author} {\bibinfo {author} {\bibfnamefont {V.}~\bibnamefont
  {Vadimov}}, \bibinfo {author} {\bibfnamefont {T.}~\bibnamefont {Hyart}},
  \bibinfo {author} {\bibfnamefont {J.~L.}\ \bibnamefont {Lado}}, \bibinfo
  {author} {\bibfnamefont {M.}~\bibnamefont {M\"ott\"onen}}, \ and\ \bibinfo
  {author} {\bibfnamefont {T.}~\bibnamefont {Ala-Nissila}},\ }\href {\doibase
  10.1103/PhysRevResearch.3.023002} {\bibfield  {journal} {\bibinfo  {journal}
  {Phys. Rev. Research}\ }\textbf {\bibinfo {volume} {3}},\ \bibinfo {pages}
  {023002} (\bibinfo {year} {2021})}\BibitemShut {NoStop}%
\bibitem [{\citenamefont {Weston}\ \emph {et~al.}(2015)\citenamefont {Weston},
  \citenamefont {Gaury},\ and\ \citenamefont {Waintal}}]{Weston2015}%
  \BibitemOpen
  \bibfield  {author} {\bibinfo {author} {\bibfnamefont {J.}~\bibnamefont
  {Weston}}, \bibinfo {author} {\bibfnamefont {B.}~\bibnamefont {Gaury}}, \
  and\ \bibinfo {author} {\bibfnamefont {X.}~\bibnamefont {Waintal}},\ }\href
  {\doibase 10.1103/PhysRevB.92.020513} {\bibfield  {journal} {\bibinfo
  {journal} {Phys. Rev. B}\ }\textbf {\bibinfo {volume} {92}},\ \bibinfo
  {pages} {020513} (\bibinfo {year} {2015})}\BibitemShut {NoStop}%
\bibitem [{\citenamefont {Francica}\ \emph {et~al.}(2016)\citenamefont
  {Francica}, \citenamefont {Apollaro}, \citenamefont {Lo~Gullo},\ and\
  \citenamefont {Plastina}}]{Francica2016}%
  \BibitemOpen
  \bibfield  {author} {\bibinfo {author} {\bibfnamefont {G.}~\bibnamefont
  {Francica}}, \bibinfo {author} {\bibfnamefont {T.~J.~G.}\ \bibnamefont
  {Apollaro}}, \bibinfo {author} {\bibfnamefont {N.}~\bibnamefont {Lo~Gullo}},
  \ and\ \bibinfo {author} {\bibfnamefont {F.}~\bibnamefont {Plastina}},\
  }\href {\doibase 10.1103/PhysRevB.94.245103} {\bibfield  {journal} {\bibinfo
  {journal} {Phys. Rev. B}\ }\textbf {\bibinfo {volume} {94}},\ \bibinfo
  {pages} {245103} (\bibinfo {year} {2016})}\BibitemShut {NoStop}%
\bibitem [{\citenamefont {Bondyopadhaya}\ and\ \citenamefont
  {Roy}(2019)}]{Bondyopadhaya2019}%
  \BibitemOpen
  \bibfield  {author} {\bibinfo {author} {\bibfnamefont {N.}~\bibnamefont
  {Bondyopadhaya}}\ and\ \bibinfo {author} {\bibfnamefont {D.}~\bibnamefont
  {Roy}},\ }\href {\doibase 10.1103/PhysRevB.99.214514} {\bibfield  {journal}
  {\bibinfo  {journal} {Phys. Rev. B}\ }\textbf {\bibinfo {volume} {99}},\
  \bibinfo {pages} {214514} (\bibinfo {year} {2019})}\BibitemShut {NoStop}%
\bibitem [{\citenamefont {Tuovinen}\ \emph
  {et~al.}(2019{\natexlab{a}})\citenamefont {Tuovinen}, \citenamefont
  {Perfetto}, \citenamefont {van Leeuwen}, \citenamefont {Stefanucci},\ and\
  \citenamefont {Sentef}}]{Tuovinen2019NJP}%
  \BibitemOpen
  \bibfield  {author} {\bibinfo {author} {\bibfnamefont {R.}~\bibnamefont
  {Tuovinen}}, \bibinfo {author} {\bibfnamefont {E.}~\bibnamefont {Perfetto}},
  \bibinfo {author} {\bibfnamefont {R.}~\bibnamefont {van Leeuwen}}, \bibinfo
  {author} {\bibfnamefont {G.}~\bibnamefont {Stefanucci}}, \ and\ \bibinfo
  {author} {\bibfnamefont {M.~A.}\ \bibnamefont {Sentef}},\ }\href {\doibase
  10.1088/1367-2630/ab4ab7} {\bibfield  {journal} {\bibinfo  {journal} {New J.
  Phys.}\ }\textbf {\bibinfo {volume} {21}},\ \bibinfo {pages} {103038}
  (\bibinfo {year} {2019}{\natexlab{a}})}\BibitemShut {NoStop}%
\bibitem [{\citenamefont {V{\"a}yrynen}\ \emph {et~al.}(2020)\citenamefont
  {V{\"a}yrynen}, \citenamefont {Pikulin},\ and\ \citenamefont
  {Lutchyn}}]{Vayrynen2020}%
  \BibitemOpen
  \bibfield  {author} {\bibinfo {author} {\bibfnamefont {J.~I.}\ \bibnamefont
  {V{\"a}yrynen}}, \bibinfo {author} {\bibfnamefont {D.~I.}\ \bibnamefont
  {Pikulin}}, \ and\ \bibinfo {author} {\bibfnamefont {R.~M.}\ \bibnamefont
  {Lutchyn}},\ }\href {https://arxiv.org/abs/2010.05963} {\bibfield  {journal}
  {\bibinfo  {journal} {arXiv:2010.05963}\ } (\bibinfo {year}
  {2020})}\BibitemShut {NoStop}%
\bibitem [{\citenamefont {Baranski}\ \emph {et~al.}(2020)\citenamefont
  {Baranski}, \citenamefont {Baranska}, \citenamefont {Zienkiewicz},
  \citenamefont {Taranko},\ and\ \citenamefont {Domanski}}]{Baranski2020}%
  \BibitemOpen
  \bibfield  {author} {\bibinfo {author} {\bibfnamefont {J.}~\bibnamefont
  {Baranski}}, \bibinfo {author} {\bibfnamefont {M.}~\bibnamefont {Baranska}},
  \bibinfo {author} {\bibfnamefont {T.}~\bibnamefont {Zienkiewicz}}, \bibinfo
  {author} {\bibfnamefont {R.}~\bibnamefont {Taranko}}, \ and\ \bibinfo
  {author} {\bibfnamefont {T.}~\bibnamefont {Domanski}},\ }\href
  {https://arxiv.org/abs/2012.03077} {\bibfield  {journal} {\bibinfo  {journal}
  {arXiv:2012.03077}\ } (\bibinfo {year} {2020})}\BibitemShut {NoStop}%
\bibitem [{\citenamefont {Vasseur}\ \emph {et~al.}(2014)\citenamefont
  {Vasseur}, \citenamefont {Dahlhaus},\ and\ \citenamefont
  {Moore}}]{Vasseur2014}%
  \BibitemOpen
  \bibfield  {author} {\bibinfo {author} {\bibfnamefont {R.}~\bibnamefont
  {Vasseur}}, \bibinfo {author} {\bibfnamefont {J.~P.}\ \bibnamefont
  {Dahlhaus}}, \ and\ \bibinfo {author} {\bibfnamefont {J.~E.}\ \bibnamefont
  {Moore}},\ }\href {\doibase 10.1103/PhysRevX.4.041007} {\bibfield  {journal}
  {\bibinfo  {journal} {Phys. Rev. X}\ }\textbf {\bibinfo {volume} {4}},\
  \bibinfo {pages} {041007} (\bibinfo {year} {2014})}\BibitemShut {NoStop}%
\bibitem [{\citenamefont {Wrze\ifmmode~\acute{s}\else \'{s}\fi{}niewski}\ and\
  \citenamefont {Weymann}(2021)}]{Wrzesniewski2021}%
  \BibitemOpen
  \bibfield  {author} {\bibinfo {author} {\bibfnamefont {K.}~\bibnamefont
  {Wrze\ifmmode~\acute{s}\else \'{s}\fi{}niewski}}\ and\ \bibinfo {author}
  {\bibfnamefont {I.}~\bibnamefont {Weymann}},\ }\href {\doibase
  10.1103/PhysRevB.103.125413} {\bibfield  {journal} {\bibinfo  {journal}
  {Phys. Rev. B}\ }\textbf {\bibinfo {volume} {103}},\ \bibinfo {pages}
  {125413} (\bibinfo {year} {2021})}\BibitemShut {NoStop}%
\bibitem [{\citenamefont {McIver}\ \emph {et~al.}(2020)\citenamefont {McIver},
  \citenamefont {Schulte}, \citenamefont {Stein}, \citenamefont {Matsuyama},
  \citenamefont {Jotzu}, \citenamefont {Meier},\ and\ \citenamefont
  {Cavalleri}}]{McIver2020}%
  \BibitemOpen
  \bibfield  {author} {\bibinfo {author} {\bibfnamefont {J.~W.}\ \bibnamefont
  {McIver}}, \bibinfo {author} {\bibfnamefont {B.}~\bibnamefont {Schulte}},
  \bibinfo {author} {\bibfnamefont {F.-U.}\ \bibnamefont {Stein}}, \bibinfo
  {author} {\bibfnamefont {T.}~\bibnamefont {Matsuyama}}, \bibinfo {author}
  {\bibfnamefont {G.}~\bibnamefont {Jotzu}}, \bibinfo {author} {\bibfnamefont
  {G.}~\bibnamefont {Meier}}, \ and\ \bibinfo {author} {\bibfnamefont
  {A.}~\bibnamefont {Cavalleri}},\ }\href {\doibase 10.1038/s41567-019-0698-y}
  {\bibfield  {journal} {\bibinfo  {journal} {Nat. Phys.}\ }\textbf {\bibinfo
  {volume} {16}},\ \bibinfo {pages} {38} (\bibinfo {year} {2020})}\BibitemShut
  {NoStop}%
\bibitem [{\citenamefont {Lee}\ \emph {et~al.}(2020)\citenamefont {Lee},
  \citenamefont {Rohwer}, \citenamefont {Sie}, \citenamefont {Zong},
  \citenamefont {Baldini}, \citenamefont {Straquadine}, \citenamefont
  {Walmsley}, \citenamefont {Gardner}, \citenamefont {Lee}, \citenamefont
  {Fisher},\ and\ \citenamefont {Gedik}}]{Lee2020}%
  \BibitemOpen
  \bibfield  {author} {\bibinfo {author} {\bibfnamefont {C.}~\bibnamefont
  {Lee}}, \bibinfo {author} {\bibfnamefont {T.}~\bibnamefont {Rohwer}},
  \bibinfo {author} {\bibfnamefont {E.~J.}\ \bibnamefont {Sie}}, \bibinfo
  {author} {\bibfnamefont {A.}~\bibnamefont {Zong}}, \bibinfo {author}
  {\bibfnamefont {E.}~\bibnamefont {Baldini}}, \bibinfo {author} {\bibfnamefont
  {J.}~\bibnamefont {Straquadine}}, \bibinfo {author} {\bibfnamefont
  {P.}~\bibnamefont {Walmsley}}, \bibinfo {author} {\bibfnamefont
  {D.}~\bibnamefont {Gardner}}, \bibinfo {author} {\bibfnamefont {Y.~S.}\
  \bibnamefont {Lee}}, \bibinfo {author} {\bibfnamefont {I.~R.}\ \bibnamefont
  {Fisher}}, \ and\ \bibinfo {author} {\bibfnamefont {N.}~\bibnamefont
  {Gedik}},\ }\href {\doibase 10.1063/1.5139556} {\bibfield  {journal}
  {\bibinfo  {journal} {Rev. Sci. Instrum}\ }\textbf {\bibinfo {volume} {91}},\
  \bibinfo {pages} {043102} (\bibinfo {year} {2020})}\BibitemShut {NoStop}%
\bibitem [{\citenamefont {Nuske}\ \emph {et~al.}(2020)\citenamefont {Nuske},
  \citenamefont {Broers}, \citenamefont {Schulte}, \citenamefont {Jotzu},
  \citenamefont {Sato}, \citenamefont {Cavalleri}, \citenamefont {Rubio},
  \citenamefont {McIver},\ and\ \citenamefont {Mathey}}]{Nuske2020}%
  \BibitemOpen
  \bibfield  {author} {\bibinfo {author} {\bibfnamefont {M.}~\bibnamefont
  {Nuske}}, \bibinfo {author} {\bibfnamefont {L.}~\bibnamefont {Broers}},
  \bibinfo {author} {\bibfnamefont {B.}~\bibnamefont {Schulte}}, \bibinfo
  {author} {\bibfnamefont {G.}~\bibnamefont {Jotzu}}, \bibinfo {author}
  {\bibfnamefont {S.~A.}\ \bibnamefont {Sato}}, \bibinfo {author}
  {\bibfnamefont {A.}~\bibnamefont {Cavalleri}}, \bibinfo {author}
  {\bibfnamefont {A.}~\bibnamefont {Rubio}}, \bibinfo {author} {\bibfnamefont
  {J.~W.}\ \bibnamefont {McIver}}, \ and\ \bibinfo {author} {\bibfnamefont
  {L.}~\bibnamefont {Mathey}},\ }\href {\doibase
  10.1103/PhysRevResearch.2.043408} {\bibfield  {journal} {\bibinfo  {journal}
  {Phys. Rev. Research}\ }\textbf {\bibinfo {volume} {2}},\ \bibinfo {pages}
  {043408} (\bibinfo {year} {2020})}\BibitemShut {NoStop}%
\bibitem [{\citenamefont {Abdo}\ \emph {et~al.}(2021)\citenamefont {Abdo},
  \citenamefont {Sheng}, \citenamefont {Rolf-Pissarczyk}, \citenamefont
  {Arnhold}, \citenamefont {Burgess}, \citenamefont {Isobe}, \citenamefont
  {Malavolti},\ and\ \citenamefont {Loth}}]{Abdo2021}%
  \BibitemOpen
  \bibfield  {author} {\bibinfo {author} {\bibfnamefont {M.}~\bibnamefont
  {Abdo}}, \bibinfo {author} {\bibfnamefont {S.}~\bibnamefont {Sheng}},
  \bibinfo {author} {\bibfnamefont {S.}~\bibnamefont {Rolf-Pissarczyk}},
  \bibinfo {author} {\bibfnamefont {L.}~\bibnamefont {Arnhold}}, \bibinfo
  {author} {\bibfnamefont {J.~A.~J.}\ \bibnamefont {Burgess}}, \bibinfo
  {author} {\bibfnamefont {M.}~\bibnamefont {Isobe}}, \bibinfo {author}
  {\bibfnamefont {L.}~\bibnamefont {Malavolti}}, \ and\ \bibinfo {author}
  {\bibfnamefont {S.}~\bibnamefont {Loth}},\ }\href {\doibase
  10.1021/acsphotonics.0c01652} {\bibfield  {journal} {\bibinfo  {journal} {ACS
  Photonics}\ }\textbf {\bibinfo {volume} {8}},\ \bibinfo {pages} {702}
  (\bibinfo {year} {2021})}\BibitemShut {NoStop}%
\bibitem [{\citenamefont {Budden}\ \emph {et~al.}(2021)\citenamefont {Budden},
  \citenamefont {Gebert}, \citenamefont {Buzzi}, \citenamefont {Jotzu},
  \citenamefont {Wang}, \citenamefont {Matsuyama}, \citenamefont {Meier},
  \citenamefont {Laplace}, \citenamefont {Pontiroli}, \citenamefont
  {Ricc{\`o}}, \citenamefont {Schlawin}, \citenamefont {Jaksch},\ and\
  \citenamefont {Cavalleri}}]{Budden2021}%
  \BibitemOpen
  \bibfield  {author} {\bibinfo {author} {\bibfnamefont {M.}~\bibnamefont
  {Budden}}, \bibinfo {author} {\bibfnamefont {T.}~\bibnamefont {Gebert}},
  \bibinfo {author} {\bibfnamefont {M.}~\bibnamefont {Buzzi}}, \bibinfo
  {author} {\bibfnamefont {G.}~\bibnamefont {Jotzu}}, \bibinfo {author}
  {\bibfnamefont {E.}~\bibnamefont {Wang}}, \bibinfo {author} {\bibfnamefont
  {T.}~\bibnamefont {Matsuyama}}, \bibinfo {author} {\bibfnamefont
  {G.}~\bibnamefont {Meier}}, \bibinfo {author} {\bibfnamefont
  {Y.}~\bibnamefont {Laplace}}, \bibinfo {author} {\bibfnamefont
  {D.}~\bibnamefont {Pontiroli}}, \bibinfo {author} {\bibfnamefont
  {M.}~\bibnamefont {Ricc{\`o}}}, \bibinfo {author} {\bibfnamefont
  {F.}~\bibnamefont {Schlawin}}, \bibinfo {author} {\bibfnamefont
  {D.}~\bibnamefont {Jaksch}}, \ and\ \bibinfo {author} {\bibfnamefont
  {A.}~\bibnamefont {Cavalleri}},\ }\href {\doibase 10.1038/s41567-020-01148-1}
  {\bibfield  {journal} {\bibinfo  {journal} {Nat. Phys.}\ }\textbf {\bibinfo
  {volume} {17}},\ \bibinfo {pages} {611} (\bibinfo {year} {2021})}\BibitemShut
  {NoStop}%
\bibitem [{\citenamefont {de~la Torre}\ \emph {et~al.}(2021)\citenamefont
  {de~la Torre}, \citenamefont {Kennes}, \citenamefont {Claassen},
  \citenamefont {Gerber}, \citenamefont {McIver},\ and\ \citenamefont
  {Sentef}}]{delaTorre2021}%
  \BibitemOpen
  \bibfield  {author} {\bibinfo {author} {\bibfnamefont {A.}~\bibnamefont
  {de~la Torre}}, \bibinfo {author} {\bibfnamefont {D.~M.}\ \bibnamefont
  {Kennes}}, \bibinfo {author} {\bibfnamefont {M.}~\bibnamefont {Claassen}},
  \bibinfo {author} {\bibfnamefont {S.}~\bibnamefont {Gerber}}, \bibinfo
  {author} {\bibfnamefont {J.~W.}\ \bibnamefont {McIver}}, \ and\ \bibinfo
  {author} {\bibfnamefont {M.~A.}\ \bibnamefont {Sentef}},\ }\href
  {https://arxiv.org/abs/2103.14888} {\bibfield  {journal} {\bibinfo  {journal}
  {arXiv:2103.14888}\ } (\bibinfo {year} {2021})}\BibitemShut {NoStop}%
\bibitem [{\citenamefont {Danielewicz}(1984)}]{Danielewicz1984}%
  \BibitemOpen
  \bibfield  {author} {\bibinfo {author} {\bibfnamefont {P.}~\bibnamefont
  {Danielewicz}},\ }\href {\doibase 10.1016/0003-4916(84)90092-7} {\bibfield
  {journal} {\bibinfo  {journal} {Ann. Phys.}\ }\textbf {\bibinfo {volume}
  {152}},\ \bibinfo {pages} {239} (\bibinfo {year} {1984})}\BibitemShut
  {NoStop}%
\bibitem [{\citenamefont {Stefanucci}\ and\ \citenamefont {van
  Leeuwen}(2013)}]{svlbook}%
  \BibitemOpen
  \bibfield  {author} {\bibinfo {author} {\bibfnamefont {G.}~\bibnamefont
  {Stefanucci}}\ and\ \bibinfo {author} {\bibfnamefont {R.}~\bibnamefont {van
  Leeuwen}},\ }\href {\doibase 10.1017/CBO9781139023979} {\emph {\bibinfo
  {title} {Nonequilibrium Many-Body Theory of Quantum Systems: A Modern
  Introduction}}}\ (\bibinfo  {publisher} {Cambridge University Press},\
  \bibinfo {year} {2013})\BibitemShut {NoStop}%
\bibitem [{\citenamefont {Balzer}\ and\ \citenamefont
  {Bonitz}(2013)}]{Balzer2013}%
  \BibitemOpen
  \bibfield  {author} {\bibinfo {author} {\bibfnamefont {K.}~\bibnamefont
  {Balzer}}\ and\ \bibinfo {author} {\bibfnamefont {M.}~\bibnamefont
  {Bonitz}},\ }\href {\doibase 10.1007/978-3-642-35082-5} {\emph {\bibinfo
  {title} {Nonequilibrium Green's Functions Approach to Inhomogeneous
  Systems}}}\ (\bibinfo  {publisher} {Springer Berlin Heidelberg},\ \bibinfo
  {year} {2013})\BibitemShut {NoStop}%
\bibitem [{\citenamefont {Schl{\"u}nzen}\ \emph {et~al.}(2020)\citenamefont
  {Schl{\"u}nzen}, \citenamefont {Hermanns}, \citenamefont {Scharnke},\ and\
  \citenamefont {Bonitz}}]{Schluenzen2020}%
  \BibitemOpen
  \bibfield  {author} {\bibinfo {author} {\bibfnamefont {N.}~\bibnamefont
  {Schl{\"u}nzen}}, \bibinfo {author} {\bibfnamefont {S.}~\bibnamefont
  {Hermanns}}, \bibinfo {author} {\bibfnamefont {M.}~\bibnamefont {Scharnke}},
  \ and\ \bibinfo {author} {\bibfnamefont {M.}~\bibnamefont {Bonitz}},\ }\href
  {\doibase 10.1088/1361-648X/ab2d32} {\bibfield  {journal} {\bibinfo
  {journal} {J. Phys. Condens. Matter}\ }\textbf {\bibinfo {volume} {32}},\
  \bibinfo {pages} {103001} (\bibinfo {year} {2020})}\BibitemShut {NoStop}%
\bibitem [{\citenamefont {Lipavsk\'y}\ \emph {et~al.}(1986)\citenamefont
  {Lipavsk\'y}, \citenamefont {\ifmmode \check{S}\else
  \v{S}\fi{}pi\ifmmode~\check{c}\else \v{c}\fi{}ka},\ and\ \citenamefont
  {Velick\'y}}]{Lipavsky1986}%
  \BibitemOpen
  \bibfield  {author} {\bibinfo {author} {\bibfnamefont {P.}~\bibnamefont
  {Lipavsk\'y}}, \bibinfo {author} {\bibfnamefont {V.}~\bibnamefont {\ifmmode
  \check{S}\else \v{S}\fi{}pi\ifmmode~\check{c}\else \v{c}\fi{}ka}}, \ and\
  \bibinfo {author} {\bibfnamefont {B.}~\bibnamefont {Velick\'y}},\ }\href
  {\doibase 10.1103/PhysRevB.34.6933} {\bibfield  {journal} {\bibinfo
  {journal} {Phys. Rev. B}\ }\textbf {\bibinfo {volume} {34}},\ \bibinfo
  {pages} {6933} (\bibinfo {year} {1986})}\BibitemShut {NoStop}%
\bibitem [{\citenamefont {{\v{S}}pi{\v{c}}ka}\ \emph
  {et~al.}(2021)\citenamefont {{\v{S}}pi{\v{c}}ka}, \citenamefont
  {Velick{\'y}},\ and\ \citenamefont {Kalvov{\'a}}}]{Spicka2021}%
  \BibitemOpen
  \bibfield  {author} {\bibinfo {author} {\bibfnamefont {V.}~\bibnamefont
  {{\v{S}}pi{\v{c}}ka}}, \bibinfo {author} {\bibfnamefont {B.}~\bibnamefont
  {Velick{\'y}}}, \ and\ \bibinfo {author} {\bibfnamefont {A.}~\bibnamefont
  {Kalvov{\'a}}},\ }\href {\doibase 10.1140/epjs/s11734-021-00109-w} {\bibfield
   {journal} {\bibinfo  {journal} {Eur. Phys. J.: Spec. Top}\ } (\bibinfo
  {year} {2021}),\ 10.1140/epjs/s11734-021-00109-w}\BibitemShut {NoStop}%
\bibitem [{\citenamefont {Fasth}\ \emph {et~al.}(2007)\citenamefont {Fasth},
  \citenamefont {Fuhrer}, \citenamefont {Samuelson}, \citenamefont {Golovach},\
  and\ \citenamefont {Loss}}]{Fasth2007}%
  \BibitemOpen
  \bibfield  {author} {\bibinfo {author} {\bibfnamefont {C.}~\bibnamefont
  {Fasth}}, \bibinfo {author} {\bibfnamefont {A.}~\bibnamefont {Fuhrer}},
  \bibinfo {author} {\bibfnamefont {L.}~\bibnamefont {Samuelson}}, \bibinfo
  {author} {\bibfnamefont {V.~N.}\ \bibnamefont {Golovach}}, \ and\ \bibinfo
  {author} {\bibfnamefont {D.}~\bibnamefont {Loss}},\ }\href {\doibase
  10.1103/PhysRevLett.98.266801} {\bibfield  {journal} {\bibinfo  {journal}
  {Phys. Rev. Lett.}\ }\textbf {\bibinfo {volume} {98}},\ \bibinfo {pages}
  {266801} (\bibinfo {year} {2007})}\BibitemShut {NoStop}%
\bibitem [{\citenamefont {van Weperen}\ \emph {et~al.}(2015)\citenamefont {van
  Weperen}, \citenamefont {Tarasinski}, \citenamefont {Eeltink}, \citenamefont
  {Pribiag}, \citenamefont {Plissard}, \citenamefont {Bakkers}, \citenamefont
  {Kouwenhoven},\ and\ \citenamefont {Wimmer}}]{vanWeperen2015}%
  \BibitemOpen
  \bibfield  {author} {\bibinfo {author} {\bibfnamefont {I.}~\bibnamefont {van
  Weperen}}, \bibinfo {author} {\bibfnamefont {B.}~\bibnamefont {Tarasinski}},
  \bibinfo {author} {\bibfnamefont {D.}~\bibnamefont {Eeltink}}, \bibinfo
  {author} {\bibfnamefont {V.~S.}\ \bibnamefont {Pribiag}}, \bibinfo {author}
  {\bibfnamefont {S.~R.}\ \bibnamefont {Plissard}}, \bibinfo {author}
  {\bibfnamefont {E.~P. A.~M.}\ \bibnamefont {Bakkers}}, \bibinfo {author}
  {\bibfnamefont {L.~P.}\ \bibnamefont {Kouwenhoven}}, \ and\ \bibinfo {author}
  {\bibfnamefont {M.}~\bibnamefont {Wimmer}},\ }\href {\doibase
  10.1103/PhysRevB.91.201413} {\bibfield  {journal} {\bibinfo  {journal} {Phys.
  Rev. B}\ }\textbf {\bibinfo {volume} {91}},\ \bibinfo {pages} {201413}
  (\bibinfo {year} {2015})}\BibitemShut {NoStop}%
\bibitem [{\citenamefont {Hermanns}\ \emph {et~al.}(2014)\citenamefont
  {Hermanns}, \citenamefont {Schl\"unzen},\ and\ \citenamefont
  {Bonitz}}]{Hermanns2014}%
  \BibitemOpen
  \bibfield  {author} {\bibinfo {author} {\bibfnamefont {S.}~\bibnamefont
  {Hermanns}}, \bibinfo {author} {\bibfnamefont {N.}~\bibnamefont
  {Schl\"unzen}}, \ and\ \bibinfo {author} {\bibfnamefont {M.}~\bibnamefont
  {Bonitz}},\ }\href {\doibase 10.1103/PhysRevB.90.125111} {\bibfield
  {journal} {\bibinfo  {journal} {Phys. Rev. B}\ }\textbf {\bibinfo {volume}
  {90}},\ \bibinfo {pages} {125111} (\bibinfo {year} {2014})}\BibitemShut
  {NoStop}%
\bibitem [{\citenamefont {Lacroix}\ \emph {et~al.}(2014)\citenamefont
  {Lacroix}, \citenamefont {Hermanns}, \citenamefont {Hinz},\ and\
  \citenamefont {Bonitz}}]{Lacroix2014}%
  \BibitemOpen
  \bibfield  {author} {\bibinfo {author} {\bibfnamefont {D.}~\bibnamefont
  {Lacroix}}, \bibinfo {author} {\bibfnamefont {S.}~\bibnamefont {Hermanns}},
  \bibinfo {author} {\bibfnamefont {C.~M.}\ \bibnamefont {Hinz}}, \ and\
  \bibinfo {author} {\bibfnamefont {M.}~\bibnamefont {Bonitz}},\ }\href
  {\doibase 10.1103/PhysRevB.90.125112} {\bibfield  {journal} {\bibinfo
  {journal} {Phys. Rev. B}\ }\textbf {\bibinfo {volume} {90}},\ \bibinfo
  {pages} {125112} (\bibinfo {year} {2014})}\BibitemShut {NoStop}%
\bibitem [{\citenamefont {Schl\"unzen}\ \emph {et~al.}(2017)\citenamefont
  {Schl\"unzen}, \citenamefont {Joost}, \citenamefont {Heidrich-Meisner},\ and\
  \citenamefont {Bonitz}}]{Schluenzen2017}%
  \BibitemOpen
  \bibfield  {author} {\bibinfo {author} {\bibfnamefont {N.}~\bibnamefont
  {Schl\"unzen}}, \bibinfo {author} {\bibfnamefont {J.-P.}\ \bibnamefont
  {Joost}}, \bibinfo {author} {\bibfnamefont {F.}~\bibnamefont
  {Heidrich-Meisner}}, \ and\ \bibinfo {author} {\bibfnamefont
  {M.}~\bibnamefont {Bonitz}},\ }\href {\doibase 10.1103/PhysRevB.95.165139}
  {\bibfield  {journal} {\bibinfo  {journal} {Phys. Rev. B}\ }\textbf {\bibinfo
  {volume} {95}},\ \bibinfo {pages} {165139} (\bibinfo {year}
  {2017})}\BibitemShut {NoStop}%
\bibitem [{\citenamefont {Rios}\ \emph {et~al.}(2011)\citenamefont {Rios},
  \citenamefont {Barker}, \citenamefont {Buchler},\ and\ \citenamefont
  {Danielewicz}}]{Rios2011}%
  \BibitemOpen
  \bibfield  {author} {\bibinfo {author} {\bibfnamefont {A.}~\bibnamefont
  {Rios}}, \bibinfo {author} {\bibfnamefont {B.}~\bibnamefont {Barker}},
  \bibinfo {author} {\bibfnamefont {M.}~\bibnamefont {Buchler}}, \ and\
  \bibinfo {author} {\bibfnamefont {P.}~\bibnamefont {Danielewicz}},\ }\href
  {\doibase 10.1016/j.aop.2010.12.009} {\bibfield  {journal} {\bibinfo
  {journal} {Ann. Phys.}\ }\textbf {\bibinfo {volume} {326}},\ \bibinfo {pages}
  {1274} (\bibinfo {year} {2011})}\BibitemShut {NoStop}%
\bibitem [{\citenamefont {Hermanns}\ \emph {et~al.}(2012)\citenamefont
  {Hermanns}, \citenamefont {Balzer},\ and\ \citenamefont
  {Bonitz}}]{Hermanns2012}%
  \BibitemOpen
  \bibfield  {author} {\bibinfo {author} {\bibfnamefont {S.}~\bibnamefont
  {Hermanns}}, \bibinfo {author} {\bibfnamefont {K.}~\bibnamefont {Balzer}}, \
  and\ \bibinfo {author} {\bibfnamefont {M.}~\bibnamefont {Bonitz}},\ }\href
  {\doibase 10.1088/0031-8949/2012/t151/014036} {\bibfield  {journal} {\bibinfo
   {journal} {Phys. Scr.}\ }\textbf {\bibinfo {volume} {T151}},\ \bibinfo
  {pages} {014036} (\bibinfo {year} {2012})}\BibitemShut {NoStop}%
\bibitem [{\citenamefont {Stan}\ \emph {et~al.}(2009)\citenamefont {Stan},
  \citenamefont {Dahlen},\ and\ \citenamefont {van Leeuwen}}]{Stan2009}%
  \BibitemOpen
  \bibfield  {author} {\bibinfo {author} {\bibfnamefont {A.}~\bibnamefont
  {Stan}}, \bibinfo {author} {\bibfnamefont {N.~E.}\ \bibnamefont {Dahlen}}, \
  and\ \bibinfo {author} {\bibfnamefont {R.}~\bibnamefont {van Leeuwen}},\
  }\href {\doibase 10.1063/1.3127247} {\bibfield  {journal} {\bibinfo
  {journal} {J. Chem. Phys.}\ }\textbf {\bibinfo {volume} {130}},\ \bibinfo
  {pages} {224101} (\bibinfo {year} {2009})}\BibitemShut {NoStop}%
\bibitem [{\citenamefont {Tuovinen}\ \emph
  {et~al.}(2019{\natexlab{b}})\citenamefont {Tuovinen}, \citenamefont
  {Gole\v{z}}, \citenamefont {Sch{\"u}ler}, \citenamefont {Werner},
  \citenamefont {Eckstein},\ and\ \citenamefont {Sentef}}]{Tuovinen2019pssb}%
  \BibitemOpen
  \bibfield  {author} {\bibinfo {author} {\bibfnamefont {R.}~\bibnamefont
  {Tuovinen}}, \bibinfo {author} {\bibfnamefont {D.}~\bibnamefont {Gole\v{z}}},
  \bibinfo {author} {\bibfnamefont {M.}~\bibnamefont {Sch{\"u}ler}}, \bibinfo
  {author} {\bibfnamefont {P.}~\bibnamefont {Werner}}, \bibinfo {author}
  {\bibfnamefont {M.}~\bibnamefont {Eckstein}}, \ and\ \bibinfo {author}
  {\bibfnamefont {M.~A.}\ \bibnamefont {Sentef}},\ }\href {\doibase
  https://doi.org/10.1002/pssb.201800469} {\bibfield  {journal} {\bibinfo
  {journal} {Phys. Status Solidi B}\ }\textbf {\bibinfo {volume} {256}},\
  \bibinfo {pages} {1800469} (\bibinfo {year}
  {2019}{\natexlab{b}})}\BibitemShut {NoStop}%
\bibitem [{\citenamefont {Ridley}\ \emph {et~al.}(2019)\citenamefont {Ridley},
  \citenamefont {Sentef},\ and\ \citenamefont {Tuovinen}}]{Ridley2019entropy}%
  \BibitemOpen
  \bibfield  {author} {\bibinfo {author} {\bibfnamefont {M.}~\bibnamefont
  {Ridley}}, \bibinfo {author} {\bibfnamefont {M.~A.}\ \bibnamefont {Sentef}},
  \ and\ \bibinfo {author} {\bibfnamefont {R.}~\bibnamefont {Tuovinen}},\
  }\href {\doibase 10.3390/e21080737} {\bibfield  {journal} {\bibinfo
  {journal} {Entropy}\ }\textbf {\bibinfo {volume} {21}},\ \bibinfo {pages}
  {737} (\bibinfo {year} {2019})}\BibitemShut {NoStop}%
\bibitem [{\citenamefont {Blackman}\ and\ \citenamefont
  {Tukey}(1959)}]{Blackman1959}%
  \BibitemOpen
  \bibfield  {author} {\bibinfo {author} {\bibfnamefont {R.~B.}\ \bibnamefont
  {Blackman}}\ and\ \bibinfo {author} {\bibfnamefont {J.~W.}\ \bibnamefont
  {Tukey}},\ }\href@noop {} {\emph {\bibinfo {title} {Particular Pairs of
  Windows: In The Measurement of Power Spectra, From the Point of View of
  Communications Engineering}}}\ (\bibinfo  {publisher} {Dover},\ \bibinfo
  {address} {New York},\ \bibinfo {year} {1959})\BibitemShut {NoStop}%
\bibitem [{\citenamefont {My\"{o}h\"{a}nen}\ \emph {et~al.}(2010)\citenamefont
  {My\"{o}h\"{a}nen}, \citenamefont {Stan}, \citenamefont {Stefanucci},\ and\
  \citenamefont {van Leeuwen}}]{Myohanen2010}%
  \BibitemOpen
  \bibfield  {author} {\bibinfo {author} {\bibfnamefont {P.}~\bibnamefont
  {My\"{o}h\"{a}nen}}, \bibinfo {author} {\bibfnamefont {A.}~\bibnamefont
  {Stan}}, \bibinfo {author} {\bibfnamefont {G.}~\bibnamefont {Stefanucci}}, \
  and\ \bibinfo {author} {\bibfnamefont {R.}~\bibnamefont {van Leeuwen}},\
  }\href {\doibase 10.1088/1742-6596/220/1/012017} {\bibfield  {journal}
  {\bibinfo  {journal} {J. Phys. Conf. Ser.}\ }\textbf {\bibinfo {volume}
  {220}},\ \bibinfo {pages} {012017} (\bibinfo {year} {2010})}\BibitemShut
  {NoStop}%
\bibitem [{\citenamefont {Ridley}\ and\ \citenamefont
  {Tuovinen}(2017)}]{Ridley2017}%
  \BibitemOpen
  \bibfield  {author} {\bibinfo {author} {\bibfnamefont {M.}~\bibnamefont
  {Ridley}}\ and\ \bibinfo {author} {\bibfnamefont {R.}~\bibnamefont
  {Tuovinen}},\ }\href {\doibase 10.1103/PhysRevB.96.195429} {\bibfield
  {journal} {\bibinfo  {journal} {Phys. Rev. B}\ }\textbf {\bibinfo {volume}
  {96}},\ \bibinfo {pages} {195429} (\bibinfo {year} {2017})}\BibitemShut
  {NoStop}%
\bibitem [{\citenamefont {Tuovinen}\ \emph
  {et~al.}(2019{\natexlab{c}})\citenamefont {Tuovinen}, \citenamefont {Sentef},
  \citenamefont {Gomes~da Rocha},\ and\ \citenamefont
  {Ferreira}}]{Tuovinen2019nano}%
  \BibitemOpen
  \bibfield  {author} {\bibinfo {author} {\bibfnamefont {R.}~\bibnamefont
  {Tuovinen}}, \bibinfo {author} {\bibfnamefont {M.~A.}\ \bibnamefont
  {Sentef}}, \bibinfo {author} {\bibfnamefont {C.}~\bibnamefont {Gomes~da
  Rocha}}, \ and\ \bibinfo {author} {\bibfnamefont {M.~S.}\ \bibnamefont
  {Ferreira}},\ }\href {\doibase 10.1039/C9NR02738F} {\bibfield  {journal}
  {\bibinfo  {journal} {Nanoscale}\ }\textbf {\bibinfo {volume} {11}},\
  \bibinfo {pages} {12296} (\bibinfo {year} {2019}{\natexlab{c}})}\BibitemShut
  {NoStop}%
\bibitem [{\citenamefont {Tuovinen}\ \emph {et~al.}(2020)\citenamefont
  {Tuovinen}, \citenamefont {Gole\ifmmode~\check{z}\else \v{z}\fi{}},
  \citenamefont {Eckstein},\ and\ \citenamefont {Sentef}}]{Tuovinen2020}%
  \BibitemOpen
  \bibfield  {author} {\bibinfo {author} {\bibfnamefont {R.}~\bibnamefont
  {Tuovinen}}, \bibinfo {author} {\bibfnamefont {D.}~\bibnamefont
  {Gole\ifmmode~\check{z}\else \v{z}\fi{}}}, \bibinfo {author} {\bibfnamefont
  {M.}~\bibnamefont {Eckstein}}, \ and\ \bibinfo {author} {\bibfnamefont
  {M.~A.}\ \bibnamefont {Sentef}},\ }\href {\doibase
  10.1103/PhysRevB.102.115157} {\bibfield  {journal} {\bibinfo  {journal}
  {Phys. Rev. B}\ }\textbf {\bibinfo {volume} {102}},\ \bibinfo {pages}
  {115157} (\bibinfo {year} {2020})}\BibitemShut {NoStop}%
\bibitem [{\citenamefont {Meir}\ and\ \citenamefont
  {Wingreen}(1992)}]{Meir1992}%
  \BibitemOpen
  \bibfield  {author} {\bibinfo {author} {\bibfnamefont {Y.}~\bibnamefont
  {Meir}}\ and\ \bibinfo {author} {\bibfnamefont {N.~S.}\ \bibnamefont
  {Wingreen}},\ }\href {\doibase 10.1103/PhysRevLett.68.2512} {\bibfield
  {journal} {\bibinfo  {journal} {Phys. Rev. Lett.}\ }\textbf {\bibinfo
  {volume} {68}},\ \bibinfo {pages} {2512} (\bibinfo {year}
  {1992})}\BibitemShut {NoStop}%
\bibitem [{\citenamefont {Zhu}\ \emph {et~al.}(2005)\citenamefont {Zhu},
  \citenamefont {Maciejko}, \citenamefont {Ji}, \citenamefont {Guo},\ and\
  \citenamefont {Wang}}]{Zhu2005}%
  \BibitemOpen
  \bibfield  {author} {\bibinfo {author} {\bibfnamefont {Y.}~\bibnamefont
  {Zhu}}, \bibinfo {author} {\bibfnamefont {J.}~\bibnamefont {Maciejko}},
  \bibinfo {author} {\bibfnamefont {T.}~\bibnamefont {Ji}}, \bibinfo {author}
  {\bibfnamefont {H.}~\bibnamefont {Guo}}, \ and\ \bibinfo {author}
  {\bibfnamefont {J.}~\bibnamefont {Wang}},\ }\href {\doibase
  10.1103/PhysRevB.71.075317} {\bibfield  {journal} {\bibinfo  {journal} {Phys.
  Rev. B}\ }\textbf {\bibinfo {volume} {71}},\ \bibinfo {pages} {075317}
  (\bibinfo {year} {2005})}\BibitemShut {NoStop}%
\bibitem [{\citenamefont {Verzijl}\ \emph {et~al.}(2013)\citenamefont
  {Verzijl}, \citenamefont {Seldenthuis},\ and\ \citenamefont
  {Thijssen}}]{Verzijl2013}%
  \BibitemOpen
  \bibfield  {author} {\bibinfo {author} {\bibfnamefont {C.~J.~O.}\
  \bibnamefont {Verzijl}}, \bibinfo {author} {\bibfnamefont {J.~S.}\
  \bibnamefont {Seldenthuis}}, \ and\ \bibinfo {author} {\bibfnamefont {J.~M.}\
  \bibnamefont {Thijssen}},\ }\href {\doibase 10.1063/1.4793259} {\bibfield
  {journal} {\bibinfo  {journal} {J. Chem. Phys.}\ }\textbf {\bibinfo {volume}
  {138}},\ \bibinfo {pages} {094102} (\bibinfo {year} {2013})}\BibitemShut
  {NoStop}%
\bibitem [{\citenamefont {Covito}\ \emph {et~al.}(2018)\citenamefont {Covito},
  \citenamefont {Eich}, \citenamefont {Tuovinen}, \citenamefont {Sentef},\ and\
  \citenamefont {Rubio}}]{Covito2018}%
  \BibitemOpen
  \bibfield  {author} {\bibinfo {author} {\bibfnamefont {F.}~\bibnamefont
  {Covito}}, \bibinfo {author} {\bibfnamefont {F.~G.}\ \bibnamefont {Eich}},
  \bibinfo {author} {\bibfnamefont {R.}~\bibnamefont {Tuovinen}}, \bibinfo
  {author} {\bibfnamefont {M.~A.}\ \bibnamefont {Sentef}}, \ and\ \bibinfo
  {author} {\bibfnamefont {A.}~\bibnamefont {Rubio}},\ }\href {\doibase
  10.1021/acs.jctc.8b00077} {\bibfield  {journal} {\bibinfo  {journal} {J.
  Chem. Theory Comput.}\ }\textbf {\bibinfo {volume} {14}},\ \bibinfo {pages}
  {2495} (\bibinfo {year} {2018})}\BibitemShut {NoStop}%
\bibitem [{\citenamefont {Cosco}\ \emph {et~al.}(2020)\citenamefont {Cosco},
  \citenamefont {Talarico}, \citenamefont {Tuovinen},\ and\ \citenamefont
  {Gullo}}]{Cosco2020}%
  \BibitemOpen
  \bibfield  {author} {\bibinfo {author} {\bibfnamefont {F.}~\bibnamefont
  {Cosco}}, \bibinfo {author} {\bibfnamefont {N.~W.}\ \bibnamefont {Talarico}},
  \bibinfo {author} {\bibfnamefont {R.}~\bibnamefont {Tuovinen}}, \ and\
  \bibinfo {author} {\bibfnamefont {N.~L.}\ \bibnamefont {Gullo}},\ }\href
  {https://arxiv.org/abs/2007.08901} {\bibfield  {journal} {\bibinfo  {journal}
  {arXiv:2007.08901}\ } (\bibinfo {year} {2020})}\BibitemShut {NoStop}%
\bibitem [{\citenamefont {Stefanucci}\ and\ \citenamefont
  {Almbladh}(2004)}]{Stefanucci2004}%
  \BibitemOpen
  \bibfield  {author} {\bibinfo {author} {\bibfnamefont {G.}~\bibnamefont
  {Stefanucci}}\ and\ \bibinfo {author} {\bibfnamefont {C.-O.}\ \bibnamefont
  {Almbladh}},\ }\href {\doibase 10.1103/PhysRevB.69.195318} {\bibfield
  {journal} {\bibinfo  {journal} {Phys. Rev. B}\ }\textbf {\bibinfo {volume}
  {69}},\ \bibinfo {pages} {195318} (\bibinfo {year} {2004})}\BibitemShut
  {NoStop}%
\bibitem [{\citenamefont {Karlsson}\ \emph {et~al.}(2018)\citenamefont
  {Karlsson}, \citenamefont {van Leeuwen}, \citenamefont {Perfetto},\ and\
  \citenamefont {Stefanucci}}]{Karlsson2018}%
  \BibitemOpen
  \bibfield  {author} {\bibinfo {author} {\bibfnamefont {D.}~\bibnamefont
  {Karlsson}}, \bibinfo {author} {\bibfnamefont {R.}~\bibnamefont {van
  Leeuwen}}, \bibinfo {author} {\bibfnamefont {E.}~\bibnamefont {Perfetto}}, \
  and\ \bibinfo {author} {\bibfnamefont {G.}~\bibnamefont {Stefanucci}},\
  }\href {\doibase 10.1103/PhysRevB.98.115148} {\bibfield  {journal} {\bibinfo
  {journal} {Phys. Rev. B}\ }\textbf {\bibinfo {volume} {98}},\ \bibinfo
  {pages} {115148} (\bibinfo {year} {2018})}\BibitemShut {NoStop}%
\bibitem [{\citenamefont {Tuovinen}\ \emph {et~al.}(2021)\citenamefont
  {Tuovinen}, \citenamefont {van Leeuwen}, \citenamefont {Perfetto},\ and\
  \citenamefont {Stefanucci}}]{Tuovinen2021}%
  \BibitemOpen
  \bibfield  {author} {\bibinfo {author} {\bibfnamefont {R.}~\bibnamefont
  {Tuovinen}}, \bibinfo {author} {\bibfnamefont {R.}~\bibnamefont {van
  Leeuwen}}, \bibinfo {author} {\bibfnamefont {E.}~\bibnamefont {Perfetto}}, \
  and\ \bibinfo {author} {\bibfnamefont {G.}~\bibnamefont {Stefanucci}},\
  }\href {\doibase 10.1063/5.0040685} {\bibfield  {journal} {\bibinfo
  {journal} {J. Chem. Phys.}\ }\textbf {\bibinfo {volume} {154}},\ \bibinfo
  {pages} {094104} (\bibinfo {year} {2021})}\BibitemShut {NoStop}%
\bibitem [{\citenamefont {Ridley}\ and\ \citenamefont
  {Tuovinen}(2018)}]{Ridley2018}%
  \BibitemOpen
  \bibfield  {author} {\bibinfo {author} {\bibfnamefont {M.}~\bibnamefont
  {Ridley}}\ and\ \bibinfo {author} {\bibfnamefont {R.}~\bibnamefont
  {Tuovinen}},\ }\href {\doibase 10.1007/s10909-018-1880-9} {\bibfield
  {journal} {\bibinfo  {journal} {J. Low Temp. Phys.}\ }\textbf {\bibinfo
  {volume} {191}},\ \bibinfo {pages} {380} (\bibinfo {year}
  {2018})}\BibitemShut {NoStop}%
\bibitem [{\citenamefont {Tuovinen}\ \emph
  {et~al.}(2019{\natexlab{d}})\citenamefont {Tuovinen}, \citenamefont
  {Covito},\ and\ \citenamefont {Sentef}}]{Tuovinen2019JCP}%
  \BibitemOpen
  \bibfield  {author} {\bibinfo {author} {\bibfnamefont {R.}~\bibnamefont
  {Tuovinen}}, \bibinfo {author} {\bibfnamefont {F.}~\bibnamefont {Covito}}, \
  and\ \bibinfo {author} {\bibfnamefont {M.~A.}\ \bibnamefont {Sentef}},\
  }\href {\doibase 10.1063/1.5121820} {\bibfield  {journal} {\bibinfo
  {journal} {J. Chem. Phys.}\ }\textbf {\bibinfo {volume} {151}},\ \bibinfo
  {pages} {174110} (\bibinfo {year} {2019}{\natexlab{d}})}\BibitemShut
  {NoStop}%
\bibitem [{\citenamefont {Schl\"unzen}\ \emph {et~al.}(2020)\citenamefont
  {Schl\"unzen}, \citenamefont {Joost},\ and\ \citenamefont
  {Bonitz}}]{Schluenzen2020PRL}%
  \BibitemOpen
  \bibfield  {author} {\bibinfo {author} {\bibfnamefont {N.}~\bibnamefont
  {Schl\"unzen}}, \bibinfo {author} {\bibfnamefont {J.-P.}\ \bibnamefont
  {Joost}}, \ and\ \bibinfo {author} {\bibfnamefont {M.}~\bibnamefont
  {Bonitz}},\ }\href {\doibase 10.1103/PhysRevLett.124.076601} {\bibfield
  {journal} {\bibinfo  {journal} {Phys. Rev. Lett.}\ }\textbf {\bibinfo
  {volume} {124}},\ \bibinfo {pages} {076601} (\bibinfo {year}
  {2020})}\BibitemShut {NoStop}%
\bibitem [{\citenamefont {Joost}\ \emph {et~al.}(2020)\citenamefont {Joost},
  \citenamefont {Schl\"unzen},\ and\ \citenamefont {Bonitz}}]{Joost2020}%
  \BibitemOpen
  \bibfield  {author} {\bibinfo {author} {\bibfnamefont {J.-P.}\ \bibnamefont
  {Joost}}, \bibinfo {author} {\bibfnamefont {N.}~\bibnamefont {Schl\"unzen}},
  \ and\ \bibinfo {author} {\bibfnamefont {M.}~\bibnamefont {Bonitz}},\ }\href
  {\doibase 10.1103/PhysRevB.101.245101} {\bibfield  {journal} {\bibinfo
  {journal} {Phys. Rev. B}\ }\textbf {\bibinfo {volume} {101}},\ \bibinfo
  {pages} {245101} (\bibinfo {year} {2020})}\BibitemShut {NoStop}%
\bibitem [{\citenamefont {Karlsson}\ \emph {et~al.}(2020)\citenamefont
  {Karlsson}, \citenamefont {van Leeuwen}, \citenamefont {Pavlyukh},
  \citenamefont {Perfetto},\ and\ \citenamefont {Stefanucci}}]{Karlsson2020}%
  \BibitemOpen
  \bibfield  {author} {\bibinfo {author} {\bibfnamefont {D.}~\bibnamefont
  {Karlsson}}, \bibinfo {author} {\bibfnamefont {R.}~\bibnamefont {van
  Leeuwen}}, \bibinfo {author} {\bibfnamefont {Y.}~\bibnamefont {Pavlyukh}},
  \bibinfo {author} {\bibfnamefont {E.}~\bibnamefont {Perfetto}}, \ and\
  \bibinfo {author} {\bibfnamefont {G.}~\bibnamefont {Stefanucci}},\ }\href
  {https://arxiv.org/abs/2006.14965} {\bibfield  {journal} {\bibinfo  {journal}
  {arXiv:2006.14965}\ } (\bibinfo {year} {2020})}\BibitemShut {NoStop}%
\bibitem [{\citenamefont {Pavlyukh}\ \emph {et~al.}(2021)\citenamefont
  {Pavlyukh}, \citenamefont {Perfetto},\ and\ \citenamefont
  {Stefanucci}}]{Pavlyukh2021}%
  \BibitemOpen
  \bibfield  {author} {\bibinfo {author} {\bibfnamefont {Y.}~\bibnamefont
  {Pavlyukh}}, \bibinfo {author} {\bibfnamefont {E.}~\bibnamefont {Perfetto}},
  \ and\ \bibinfo {author} {\bibfnamefont {G.}~\bibnamefont {Stefanucci}},\
  }\href {https://arxiv.org/abs/2103.11932} {\bibfield  {journal} {\bibinfo
  {journal} {arXiv:2103.11932}\ } (\bibinfo {year} {2021})}\BibitemShut
  {NoStop}%
\bibitem [{\citenamefont {Zhang}\ \emph {et~al.}(2020)\citenamefont {Zhang},
  \citenamefont {L\"u},\ and\ \citenamefont {Wang}}]{Zhang2020}%
  \BibitemOpen
  \bibfield  {author} {\bibinfo {author} {\bibfnamefont {Z.-Q.}\ \bibnamefont
  {Zhang}}, \bibinfo {author} {\bibfnamefont {J.-T.}\ \bibnamefont {L\"u}}, \
  and\ \bibinfo {author} {\bibfnamefont {J.-S.}\ \bibnamefont {Wang}},\ }\href
  {\doibase 10.1103/PhysRevB.101.161406} {\bibfield  {journal} {\bibinfo
  {journal} {Phys. Rev. B}\ }\textbf {\bibinfo {volume} {101}},\ \bibinfo
  {pages} {161406} (\bibinfo {year} {2020})}\BibitemShut {NoStop}%
\bibitem [{\citenamefont {Ridley}\ \emph {et~al.}(2021)\citenamefont {Ridley},
  \citenamefont {Kantorovich}, \citenamefont {van Leeuwen},\ and\ \citenamefont
  {Tuovinen}}]{Ridley2021}%
  \BibitemOpen
  \bibfield  {author} {\bibinfo {author} {\bibfnamefont {M.}~\bibnamefont
  {Ridley}}, \bibinfo {author} {\bibfnamefont {L.}~\bibnamefont {Kantorovich}},
  \bibinfo {author} {\bibfnamefont {R.}~\bibnamefont {van Leeuwen}}, \ and\
  \bibinfo {author} {\bibfnamefont {R.}~\bibnamefont {Tuovinen}},\ }\href
  {\doibase 10.1103/PhysRevB.103.115439} {\bibfield  {journal} {\bibinfo
  {journal} {Phys. Rev. B}\ }\textbf {\bibinfo {volume} {103}},\ \bibinfo
  {pages} {115439} (\bibinfo {year} {2021})}\BibitemShut {NoStop}%
\end{thebibliography}
\end{document}